\documentclass[12pt,a4]{iopart}
\makeatletter\@namedef{ver@amsmath.sty}{}\makeatother
\usepackage{amssymb}
\usepackage{amsmath}
\usepackage{setstack}
\usepackage{graphicx}
\usepackage{tikz}
\usepackage{iopams}
\usetikzlibrary{shapes.geometric,calc}

\usetikzlibrary{shapes.geometric,calc}

\begin{document}	
	

\title[Characterization of Surface-Plasmon Polaritons at Lossy Interfaces]{Characterization of Surface-Plasmon Polaritons at Lossy Interfaces}

\author{Nafiseh Sang-Nourpour$^{1, 2, 3}$, Benjamin R. Lavoie$^{4, 2}$, R.~Kheradmand$^{1}$, M. Rezaei$^{5}$ and Barry C. Sanders$^{2, 6, 7, 8, 9}$}

\address{$^{1}$Photonics Group, Research Institute for Applied Physics and Astronomy, University of Tabriz 51665-163, Iran}
\address{$^{2}$Institute for Quantum Science and Technology, University of Calgary, Alberta T2N~1N4, Canada}
\address{$^{3}$Photonics Group, Aras International Campus, University of Tabriz 51666-16471, Iran}
\address{$^{4}$Department of Electrical and Computer Engineering, University of Calgary, Alberta T2N~1N4, Canada}
\address{$^{5}$Department of Theoretical Physics, Faculty of Physics, University of Tabriz 51664, Iran}
\address{$^{6}$Hefei National Laboratory for Physical Sciences at the Microscale,
University of Science and Technology of China, Hefei, Anhui 230026, China}
\address{$^{7}$Shanghai Branch, CAS Center for Excellence and Synergetic Innovation Center in Quantum Information and Quantum Physics, University of Science and Technology of China, Shanghai 201315, China}
\address{$^{8}$Program in Quantum Information Science, Canadian Institute for Advanced Research, Toronto, Ontario M5G~1Z8, Canada}
\address{$^{9}$Institute for Quantum Information and Matter, California Institute of Technology, Pasadena, California 91125, USA}

\ead{nafiseh.sangnourpour@ucalgary.ca}

\begin{abstract}
We characterize surface-plasmon polaritons at lossy planar interfaces between one dispersive and one nondispersive linear isotropic homogeneous media, i.e., materials or metamaterials.
Specifically we solve Maxwell's equations to obtain strict bounds for the permittivity and permeability of these media such that satisfying these bounds implies surface-plasmon polaritons successfully propagate at the interface,
and violation of the bounds impedes propagation,
i.e., the field delocalizes from the surface into the bulk.
Our characterization of surface-plasmon polaritons is valuable for checking viability of a proposed application,
and, as an example,
we employ our method to falsify a previous prediction that surface-plasmon propagation
through a surface of a double-negative refractive index medium occurs
for any permittivity and permeability;
instead we show that propagation can occur only for certain medium parameters.
\end{abstract}

\vspace{2pc}
\noindent{\it Keywords}: surface-plasmon polariton, lossy planar interface, characteristic equation

\section{Introduction}

Surface-plasmon polaritons (SPPs) are electromagnetic (EM) excitations that propagate along the interface,
and are localized to the interface,
between a conductor or exotic medium,
such as
a double-negative index metamaterial,
and a dielectric~\cite{Kamli2011, Pitarke2007, Maier2005}.
Proper characterization of SPPs is important for numerous applications,
ranging from microscopy, lithography and bio-sensing to waveguides beyond the diffraction limit~\cite{Verma2016, Gramotnev2010, Pitarke2007, Barnes2003}.
In addition SPP applications in nanophotonics and optical circuits require proper characterization to exploit their benefits.
Properties of SPPs are characterized by mathematically relating wavenumber to propagation coefficient,
i.e., a dispersion relation,
and this characteristic is obtained from eigenvalues of the electromagnetic wave equation subject to pertinent boundary conditions.
The eigenfunctions are modes,
but not all modes are propagating SPPs
as some modes delocalize from the surface
to having significant support in the bulk~\cite{Warmbier2012}.

Here we introduce strict bounds on the
real and imaginary parts of the squared propagation coefficient to decide whether a given mode is a propagating SPP or not.
Previous studies of SPPs have either focused on lossless interfaces~\cite{Tao2009, Sounas2009, Darmanyan2003}
or neglect the effects of losses on the characteristics of SPPs and consequently employ approximate but erroneous expressions for wavenumber~\cite{Kamli2011,Zeller2011,Zayats2005,Zhang2005,Barnes2003}.
Some previous works have included loss
but focus only on electric responses, 
which is pertinent for metals
but ignore the magnetic response~\cite{Derrien2016,Nguyen2014,Norrman2013,Zhang2010}.
Our results,
which are restricted to the quite general case of linear,
homogenous, isotropic (LHI) materials meeting at planar interfaces,
include all these previous results as special cases and corrects previous errors due to ignoring losses.

Furthermore we confirm the consistency of the bounds for our characteristic equations by showing that the intensity concentration of modes is localized to the surface for the squared propagation coefficient satisfying both bounds, and the Poynting vector has a large component along the propagation direction compared to the component perpendicular to the surface.
For modes that do not satisfy the bounds,
we show numerically that the Poynting vector has a larger component perpendicular to the surface compared to the component along the propagation direction
and the energy is delocalized from the surface.

SPPs are vector fields
with components conveniently labeled transverse electric (TE) and transverse magnetic (TM) SPPs.
Existence of TM SPPs
is predicated on the electric susceptibility changing sign across the interface~\cite{Zayats2005, Darmanyan2003},
Whereas, for the TE SPP,
magnetic susceptibility must change sign across the interface~\cite{Darmanyan2003}.
Lossy double-negative-index media
have been shown to support both TE and TM SPPs
for any permittivity and permeability~\cite{Kamli2011},
but here we analyze this claim carefully and show contrariwise that TE and TM SPPs
are supported only for specific values of permittivity and permeability according to our characteristic relations.

In the following we present background of SPPs
and their characterization in Sec.~\ref{sec:background},
detailed analysis in Sec.~\ref{sec:analysis}
and examples in Sec.~\ref{sec:examples}.
We discuss our results in Sec.~\ref{sec:discussion}
and conclude in Sec.~\ref{sec:conclusions}.

\section{Background}
\label{sec:background}
This section provides the background required to characterize SPP propagation at planar interfaces between dispersive and non-dispersive LHI materials.
We treat metamaterials and metals within the framework of dispersive LHI materials at the planar interface, characterized by frequency-dependent electric permittivity $\varepsilon(\omega)$ and magnetic permeability $\mu(\omega)$. 
We assume that material dispersion on the dielectric side of the interface is minimal over the spectral range of interest and refer to these materials as being nondispersive.

We employ the Drude-Lorentz model to construct the permittivity and permeability of dispersive LHI materials~\cite{Lavoie2012,Woodley2010}.
We choose the Drude-Lorentz model for its
proper microscopic description of material responses to incident EM fields.
For a dispersive LHI material with resonance frequency $\omega_{0}$, oscillation strength $F$ and damping constant $\Gamma$, the Drude-Lorentz model yields 
\begin{eqnarray}
\label{eq:permmodele}
	\frac{\varepsilon(\omega)}{\varepsilon_0}
		=&\varepsilon_{\mathrm{b}}+\frac{F_{\mathrm e} \omega_{\mathrm{e}}^2}{\omega_{0_{\mathrm{e}}}^2-\omega^2
			+\mathrm{i}\Gamma_{\mathrm{e}} \omega},\\
	\frac{\mu(\omega)}{\mu_0}
		=&\mu_{\mathrm{b}}+\frac{F_{\mathrm m} \omega_{\mathrm m}^2}{\omega_{0_{\mathrm{m}}}^2-\omega^2
			+\mathrm{i}\Gamma_{\mathrm{m}} \omega}
\label{eq:permmodel}.
\end{eqnarray}
Here
$\varepsilon_0$ ($\mu_0$) and $\varepsilon_{\mathrm{b}}$ ($\mu_{\mathrm{b}}$) are vacuum and background permittivity (permeability), respectively.
Subscripts ${}_\mathrm{e}$ and ${}_\mathrm{m}$ refer to the electric and magnetic parameters,
respectively, and
the angular frequency of the incident EM field is $\omega$.

The magnetic response of metamaterials, which are artificially engineered materials made from multiple elements such as metals and dielectrics~\cite{Shalaev2007}, is characterized by a modified Drude-Lorentz model~\cite{Lavoie2012,Penciu2010}.
This modified model can be explained using the equivalent resistor-inductor-capacitor circuit model for the metamaterial structure~\cite{Penciu2010}.
The effective magnetic permeability is
\begin{equation}
     {	\frac{\mu(\omega)}{\mu_0}=\mu_{\mathrm{b}}+\frac{F_{\mathrm m} \omega^2}{\omega_{0_{\mathrm{m}}}^2-\omega^2+\mathrm{i}\Gamma_{\mathrm{m}} \omega}},
	\label{eq:perm}
\end{equation}
where $\omega_{\mathrm m}$ in (\ref{eq:permmodel}) has been replaced by $\omega$.
The effective electric permittivity~(\ref{eq:permmodele}) for metamaterials is obtained for $\omega_{0_{\mathrm{e}}}=0$ and $F_{\mathrm e}=1$, corresponding to electrical charged particles being free and no contributions from bound charges.

Metals are an important case of dispersive LHI materials and here we discuss both the familiar form of electric metals and introduce the notion of effective magnetic metals.
Electric metals correspond to constant permeability and frequency-dependent permittivity (\ref{eq:permmodele}) with $\omega_{0_{\mathrm{e}}}=0$, $F_{\mathrm e}=1$.
For convenience,
and to aid direct comparison of results, we henceforth fix {$\Gamma_{\mathrm e}$ to be the same for all dispersive materials.
Dispersive LHI materials with constant permittivity and frequency-dependent permeability, which we call effective magnetic metals, have permeability corresponding to Eq.~(\ref{eq:permmodel}) with $F_{\mathrm m}=1$.

We employ the term effective magnetic metals, as we are fixing $\omega_{0_{\mathrm{m}}}$ to be a non-zero value (while holding permittivity constant) similar to what would be expected from simulating a magnetic metal with a metamaterial. As these materials do not have free magnetic charges, thus only approximate a magnetic metal. On the other hand choosing this approximate magnetic metal allows us to explore another distinct permittivity/permeability combination.
As with~$\Gamma_{\mathrm e}$, we fix~$\Gamma_{\mathrm m}$ to be the same for the dispersive materials throughout this paper.

Although magnetic metals do not exist in nature, effective magnetic metals could be manifested as metamaterials.
One phenomenon
that is characteristic of metals is the existence of SPPs, which can also exist for metamaterials.
Electric metals only support TM, and not TE, SPPs because of the requirement that the field propagation is parallel to the interface~\cite{Maier2007}.
Similarly, an effective magnetic metal has the property that only the TE mode is supported because the magnetic field propagates parallel to the interface.
These are necessary, but not sufficient, conditions for the existence of SPPs, and our aim is to have a full characterization of SPPs at interfaces involving either electric or effective magnetic metals.

\begin{figure}
\centering
\begin{tikzpicture}
  
 \begin{scope}[shift = {(-3,-.2)}]
 
\pgfmathsetmacro{\cubex}{5.8}
\pgfmathsetmacro{\cubey}{1.2}
\pgfmathsetmacro{\cubez}{.0}

\draw[black,fill=gray!20] (3.6,-1.7,0) -- ++(-\cubex,0,0) -- ++(0,-\cubey,0) -- ++(\cubex,0,0) -- cycle;
\draw[black,fill=gray!20] (3.6,-1.7,0) -- ++(0,0,-\cubez) -- ++(0,-\cubey,0) -- ++(0,0,\cubez) -- cycle;
\draw[black,fill=gray!20] (3.6,-1.7,0) -- ++(-\cubex,0,0) -- ++(0,0,-\cubez) -- ++(\cubex,0,0) -- cycle;

\draw[black,fill=gray!0] (3.6,-0.5,0) -- ++(-\cubex,0,0) -- ++(0,-\cubey,0) -- ++(\cubex,0,0) -- cycle;
\draw[black,fill=gray!0] (3.6,-0.5,0) -- ++(0,0,-\cubez) -- ++(0,-\cubey,0) -- ++(0,0,\cubez) -- cycle;
\draw[black,fill=gray!0] (3.6,-0.5,0) -- ++(-\cubex,0,0) -- ++(0,0,-\cubez) -- ++(\cubex,0,0) -- cycle;

    \node [below] at (3.1,-.6) {\large $2$};
   \node [below] at (3.1,-1.78) {\large $1$};

\draw [->,thick] (-2.17,-1.68) -- (-2.17,-.8);
\draw [->,thick] (-2.2,-1.68) -- (-.05,-1.68);

    \node [below] at (-1.95,-.45) {\large $x$};
   \node [below] at (-.05,-1.75) {\large $z$};
   
  \draw [->,thick] (1.2,-1.58) -- (1.2,-.95);
\draw [->,thick] (1.2,-1.58) -- (2.3,-1.58); 
   
    \node [below] at (2.2,-.9) {\large $\beta$};
   \node [below] at (.9,-.5) {\large $\gamma_{2}$};
   
\draw [very thick,dashed] (-2.2,-1.7) parabola (-.5,-.9);
\draw [very thick,dashed] (-2.2,-1.7) parabola (-.5,-2.9);

\node [below] at (-1.15,-.48) {$|E^2_z|$};
   
\end{scope}
   \end{tikzpicture}
   \caption{\label{fig:interface} Planar interface between two materials.
   Region 2 is for a lossy dispersive LHI material and region 1 is for a lossless nondispersive LHI material.
The propagation direction of SPPs is along the~$z$ axis.
Dashed lines represent the behaviour of electric field intensity for TM SPPs at the interface.
The field intensity decreases exponentially with the distance from the interface.}
\end{figure}
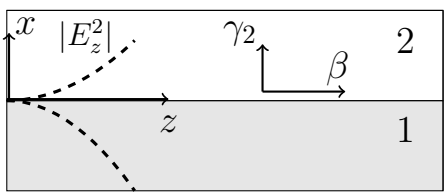
   
The characteristics of SPPs at the interfaces of LHI materials depend on the EM susceptibilities of materials at the interface,
the mode propagation coefficient and the wavenumber~\cite{Landau1984}.
The wavenumber~\cite{Lavoie2012}
\begin{eqnarray}
	\gamma_{j}(\omega)=\gamma'_{j}(\omega)+\mathrm{i} \gamma''_{j}(\omega)
		=\sqrt{\beta^2(\omega)-\omega^2 \mu_j(\omega) \varepsilon_j(\omega)}
\label{eq:gamma}
\end{eqnarray}
determines the modal behaviour in the transverse direction
($x$ direction as shown in Fig.~\ref{fig:interface})
perpendicular to propagation,
where $\gamma'_{j}$ and $\gamma''_{j}$ are real and $j=1, 2$ indicate the nondispersive and dispersive LHI materials, respectively.
The frequency-dependent propagation coefficient along the propagation direction ($z$ direction as shown in Fig.~\ref{fig:interface}) of TM SPPs for the LHI materials is~\cite{Kamli2011}
\begin{eqnarray}
\beta(\omega)=\beta'(\omega)+\mathrm{i} \beta''(\omega)=k_0 \sqrt{\varepsilon_1(\omega) \varepsilon_2(\omega) \frac{\mu_1(\omega) \varepsilon_2(\omega)-\mu_2(\omega) \varepsilon_1(\omega)}{\varepsilon_2^2(\omega)-\varepsilon_1^2(\omega)}},
\label{eq:dispersion}
\end{eqnarray}
where $k_0=\omega/c$
and
 \begin{equation}
\label{betaconditions}
         \beta'(\omega)>0, \ \beta''(\omega)>0	
\end{equation}
for forward propagation and gainless interfaces, respectively.

To characterize
SPPs we determine bounds for the squared complex propagation coefficient ($\beta^2(\omega)$) at the interface,
which is easier than working with~$\beta(\omega)$
because of the square root appearing in Eq.~(\ref{eq:dispersion}).
The real part of the squared propagation coefficient~$\beta^2(\omega)$ for TM SPPs is
\begin{eqnarray}
	\mathrm{Re}\left(\beta^2\right)
=&
k^2_0 \frac{1}{\varepsilon'^{4}_1+2\varepsilon'^2_1\left(-\varepsilon'^2_2+\varepsilon''^2_2\right)+\left(\varepsilon'^2_2+\varepsilon''^2_2\right)^2}\nonumber\\ 
&\times \Big( \varepsilon'_1 \left[ \mu'_1 \varepsilon'^2_1\left(-\varepsilon'^2_2+\varepsilon''^2_2 \right)+\mu'_1\left(\varepsilon'^2_2+\varepsilon''^2_2\right)\right.^2\nonumber\\
&\left.- \varepsilon'_1 \left(\varepsilon'^2_2+\varepsilon''^2_2\right)\left(\varepsilon'_2 \mu'_2+\varepsilon''_2 \mu''_2\right) \right]\nonumber\\
&+ \varepsilon'^{3}_1\left(\varepsilon'_2 \mu'_2-\varepsilon''_2 \mu''_2\right)
\Big)
\label{betareal}
\end{eqnarray}
and the imaginary part has the form

\begin{eqnarray}
	\mathrm{Im}\left(\beta^2\right)
=&
k^2_0 \frac{1}{\varepsilon'^{4}_1+2\varepsilon'^2_1\left(-\varepsilon'^2_2+\varepsilon''^2_2\right)+\left(\varepsilon'^2_2+\varepsilon''^2_2\right)^2}\nonumber\\ 
&\times \varepsilon'^2_1 \Big( \varepsilon''_{2} \left[-2 \mu'_1 \varepsilon'_1 \varepsilon'_{2}+ \left(\varepsilon'^2_1+\varepsilon'^2_2+\varepsilon''^2_2\right) \mu'_2 \right]\nonumber\\
&-\varepsilon'_2\left(-\varepsilon'^2_1+\varepsilon'^2_2+\varepsilon''^2_2\right) \mu''_2 \Big).
\label{betaim}
\end{eqnarray}
The propagation coefficient of TE SPPs is obtained by exchanging the EM parameters, namely applying the duality transformations to the expressions for TM modes.
With this background we are now equipped to present our method for deriving characteristic equations of SPPs.

\section{Analysis}
\label{sec:analysis}
In this section
we characterize SPPs at planar interfaces between lossy dispersive LHI materials and lossless nondispersive LHI materials.
Our approach is to apply the conditions on the wavenumber from the lossless to the lossy case and derive bounds on the square of the complex propagation coefficient, namely~$\beta^2(\omega)$, for lossy interfaces.
From these bounds on~$\beta^2(\omega)$ we construct strict bounds on~$\epsilon(\omega)$ and~$\mu(\omega)$,
such that satisfying bounds implies SPPs successfully propagates at the interface.

We establish bounds on $\beta^2$ by starting from the conditions on the wavenumber for the lossless case
and use these conditions to define what constitutes an SPP for lossy interfaces.
We suppress $(\omega)$ in frequency-dependent parameters for readability.
We start with lossless materials at the interface to have clear intuition on the problem.
For lossless materials, the wavenumber in Eq.~(\ref{eq:gamma}) is a real quantity, which requires
\begin{equation}
\beta^2>k^{2}_0 \varepsilon_1 \mu_1.
\label{bonbeta}
\end{equation}
If a small amount of loss is present in the dispersive material (material 2) near the interface (we are generalizing to an interface between one lossy and one lossless material and consider material 1 to be lossless), the wavenumber and propagation coefficient are complex quantities.
The real and imaginary parts of the wavenumber are then of the form
\begin{eqnarray}
	\gamma'_1
		=&\frac{1}{\sqrt{2}}\sqrt {\sqrt {\delta^{2}+\left(\left(\beta^2\right)''\right)^{2}}+\delta}
\label{gammar}
\end{eqnarray}
and
\begin{eqnarray}
\gamma''_1=&\frac{\mathrm{sgn}\left(\left(\beta^2\right)''\right)}{\sqrt2} \sqrt {\sqrt {\delta^{2}+\left(\left(\beta^2\right)''\right)^{2}}-\delta}
\label{gammai}
\end{eqnarray}
for
\begin{equation}
	\delta:=\left(\beta^2\right)'-k^2_0\varepsilon_1 \mu_1.
\end{equation}
In the presence of small loss,
SPPs should be nearly indistinguishable from the lossless case,
which implies that $\gamma'_1 \approx \gamma_1$ and $\gamma''_1$ should be small.

If $\delta$ is negative,
the real and imaginary parts of the wavenumber in (\ref{gammar}) and (\ref{gammai}) exchange.
This exchange is however not acceptable as $\gamma'_1$ does not behave like $\gamma_1$
in the lossless case and $\gamma''_1$ is not be small.
Therefore,
we define SPPs in the lossy case to be solutions that satisfy the condition:
\begin{equation}
	\delta>0. 
\label{brlimit}
\end{equation}
Higher values of loss in the materials give the same result as small-loss materials, as the limiting cases should hold for materials with higher losses.
Note that using Eq.~(\ref{betaconditions}) and $\left(\beta^{2}\right)''=2\beta' \beta''$ yields
\begin{equation}
	\left(\beta^2\right)''>0;
\label{bilimit}
\end{equation}
therefore,
$\mathrm{sgn}\left(\left(\beta^2\right)''\right)=1$ in (\ref{gammai}).
Equations~(\ref{brlimit}) and~(\ref{bilimit}) are bounds on the real and imaginary parts of $\beta^2$, from which we characterize SPPs.

The bounds on $\beta^2$ also apply to the bounds on the complex wavenumber.
\begin{equation}
\label{eq:gamma1}
	\left(\gamma'_1+\mathrm{i} \gamma''_1\right)^{2}=\left(\beta^{2}\right)'-k^{2}_0 \varepsilon_1 \mu_1+\mathrm{i} \left(\beta^{2}\right)''
\end{equation}
for a dielectric at the interface.
Therefore,
\begin{eqnarray}
\label{condition1}
&\left(\gamma'_1-\gamma''_1\right)\left(\gamma'_1+\gamma''_1\right)=\left(\beta^{2}\right)'-k^{2}_0 \varepsilon_1 \mu_1,\\
&2\gamma'_1 \gamma''_1=2\beta' \beta''.
\label{condition2}
\end{eqnarray}
From Eqs.~(\ref{condition1}) and (\ref{condition2}) we realize that $\vert\gamma'\vert>\vert \gamma'' \vert$ as we require (\ref{brlimit}) and (\ref{bilimit}) to hold.
This inequality is physically realizable since SPPs decay exponentially as a function of distance from the interface, which implies the mathematical condition $\vert \gamma'_1 \vert> \vert \gamma''_1 \vert$ for SPPs~\cite{Lavoie2012}.
Moreover, from Eqs.~(\ref{bilimit}) and (\ref{condition2}) we realize that $\gamma'$ and $\gamma''$ must have the same sign. As long as $\gamma'$ and $\gamma''$ are either both positive or both negative, the conditions in Eqs.~(\ref{condition1}) and (\ref{condition2}) are equivalent and consistent with (\ref{brlimit}) and (\ref{bilimit}).

The nondispersive LHI materials have permittivity $\varepsilon_1=\varepsilon'_1$ and permeability $\mu_1=\mu'_1=1$.
The dispersive LHI materials can be double-negative,~$\mu$-negative,~$\varepsilon$-negative, or double-positive.
By considering Eqs.~(\ref{bilimit}) and (\ref {brlimit}) and employing Eqs.~(\ref{betareal}) and (\ref{betaim}), 
the characteristic equations of TM SPPs at lossy dispersive LHI material interfaces with nondispersive LHI materials are
\begin{eqnarray}
0<&\varepsilon''_2<\varepsilon'_1,\quad
\varepsilon'_2<-\sqrt{\varepsilon'^2_1-\varepsilon''^2_2},\
\mu''_2\ge 0,\nonumber\\
\mu'_2&>
\frac{1}{-\varepsilon'^{2}_1 \varepsilon'_2+\varepsilon'^{3}_2+ \varepsilon'_2 \varepsilon''^2_2}\nonumber\\
&\times \Big(\varepsilon'_1 \mu'_1 \left(-\varepsilon'^2_1+\varepsilon'^2_2-\varepsilon''^{2}_{2}\right)
+\varepsilon''_{2}\mu''_{2}\left(-\varepsilon'^2_1-\varepsilon'^2_{2}-\varepsilon''^2_{2} \right)\Big).
\label{C}
\end{eqnarray}
Equation~(\ref{C}) shows that the characteristics of SPPs do not just depend on the signs of permittivity and permeability at the interface but additionally depends on the relative values of their real and imaginary parts.
We call these criteria the characteristic equations of SPPs.

The characteristic equation for TE SPPs are obtained by applying the duality transformations, i.e., exchanging the permittivity and permeability of the material.
The characteristic equations in (\ref{C}) and their dual expressions
characterize the propagation of TM and TE SPPs
at LHI planar interfaces between one lossy and one lossless material.
These equations are quite general in that the dispersive LHI materials at the interface can have both negative and positive $\varepsilon'$ and $\mu'$.
To verify the consistency of our characteristic equations, we show that the intensity concentration of modes is localized to the surface and the Poynting vector has a large component along the propagation direction compared to the component perpendicular to the surface.

To specify the applications of our characteristic equations, we first present examples of interfaces with a hypothetical dispersive LHI materials having double-negative refractive index, $\mu$-negative, $\varepsilon$-negative and double-positive refractive index.
We then treat metamaterials and metals as applicable examples of dispersive LHI materials at the interface.

\section{Examples}
\label{sec:examples}
In this section we demonstrate the applicability of our characteristic equations to existing interfaces.
We present some examples of the lossy planar interfaces of dispersive LHI materials with lossless nondispersive LHI materials.
We first present an example of a hypothetical material, having double-negative, double positive and single-negative refractive indices over different frequency ranges, forming a planar interface with air.
This is an example that is not realized in nature, but serves to clarify the characteristic equations.

We then employ planar interfaces of metamaterials, electric metals and effective magnetic metals with air as realistic examples of materials at the interface.
To have a clear understanding of SPP characteristics and to simplify the characteristic equations, we consider air (with $\varepsilon_1\approx1$ and $\mu_1\approx1$) as the lossless nondispersive LHI material at the interface, and we assume that air is dispersion-free
 over the spectral range of interest.

The characteristic equations in (\ref{C}) take a simplified form for dispersive LHI material interfaces with air:
\begin{eqnarray}
&0<\varepsilon''_2<1,\
\varepsilon'_2<-\sqrt{1-\varepsilon''^2_2},\
\mu''_2>0,\nonumber\\
& \mu'_2>\frac{-1+\varepsilon'^2_2-\varepsilon''^2_2-\varepsilon''_2 \mu''_2-\varepsilon'^2_2 \varepsilon''_2 \mu''_2-\varepsilon''^3_2 \mu''_2}{-\varepsilon'_2+\varepsilon'^{3}_2+\varepsilon'_2 \varepsilon''^2_2}.
\label{Ca}
\end{eqnarray}
At interfaces between electric metals ($\mu_2=1$) and air, the characteristic equations reduce to
\begin{eqnarray}
 \varepsilon'_2<-\sqrt{1-\varepsilon''_2},\ \ \  
 0<\varepsilon''_2<1
 \label{emetal}
\end{eqnarray}
 for TM SPPs, and TE SPPs are unsupported.
In the dual case, the characteristic equations of TE SPPs in effective magnetic metals ($\varepsilon_2=1$) is
\begin{eqnarray}
\mu'_2<-\sqrt{1-\mu''_2},\ \ \
0<\mu''_2<1,
 \label{mmetal}
\end{eqnarray}
where TM SPPs are unsupported.
With these information, we now investigate the behaviour of SPPs at the interfaces of the example materials at the interfaces.

\subsection{Interfaces of Generalized Lossy LHI Material With Air}
As the first example, we employ a generalized hypothetical dispersive LHI material interface with air.
For completeness, we choose the hypothetical dispersive LHI material that exhibits double-negative,~$\mu$-negative,~$\varepsilon$-negative and double-positive EM responses.
We investigate the behaviour of field intensity and Poynting vector at the interfaces to verify our characteristic equations.
The behaviour of the
dispersive LHI material's permittivity and permeability, field intensity, Poynting vector, $\left(\beta^2\right)'/k^{2}_0 \varepsilon_1 \mu_1$ and $\left(\beta^2\right)''$ for TM and TE SPPs at a lossy
dispersive LHI material interface with air are shown in Fig.~\ref{fig:skindepth},
which we now explain.
\begin{figure}
\centering
\includegraphics[width=.36 \columnwidth]{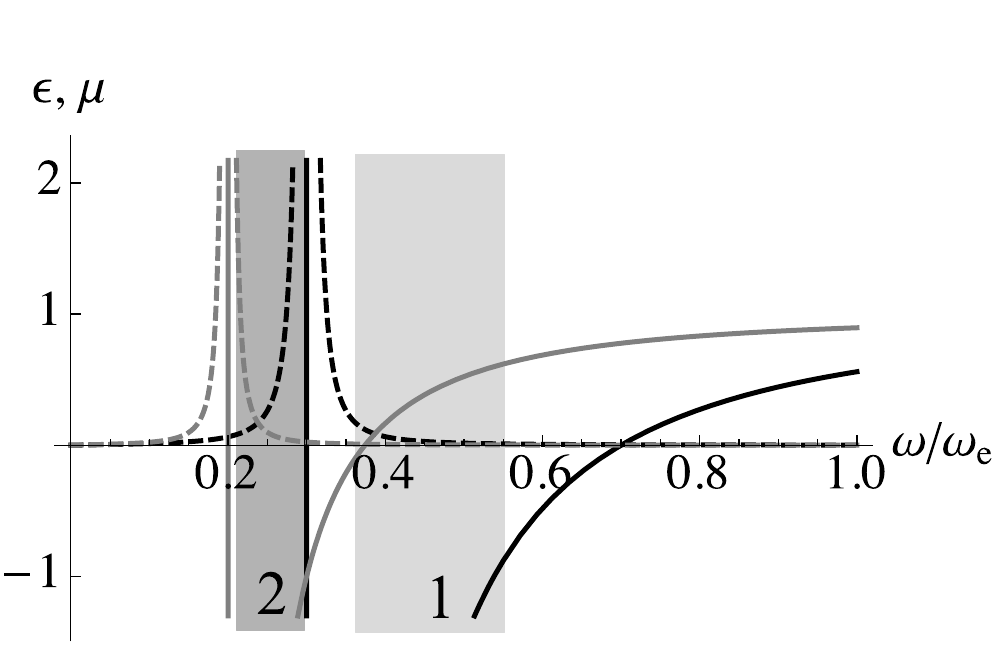}(a)\\ 
\includegraphics[width=.29 \columnwidth]{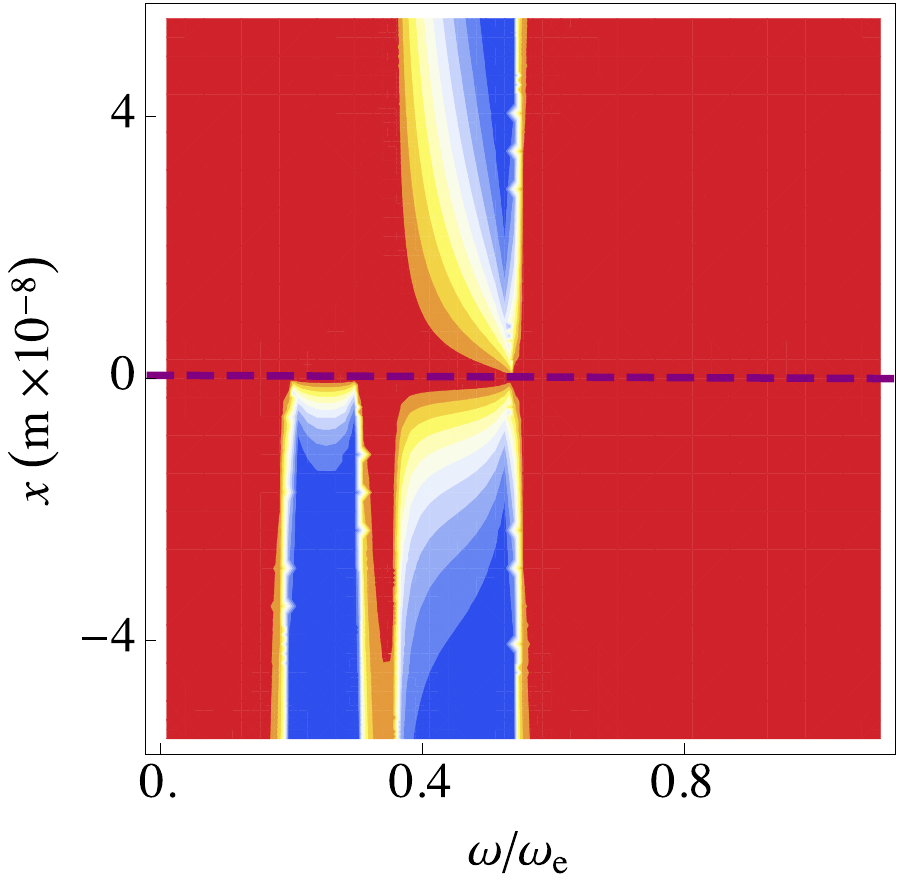}
\includegraphics[width=.042 \columnwidth]{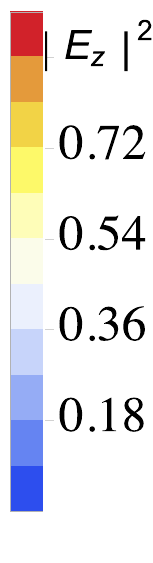}(b)
\includegraphics[width=.29 \columnwidth]{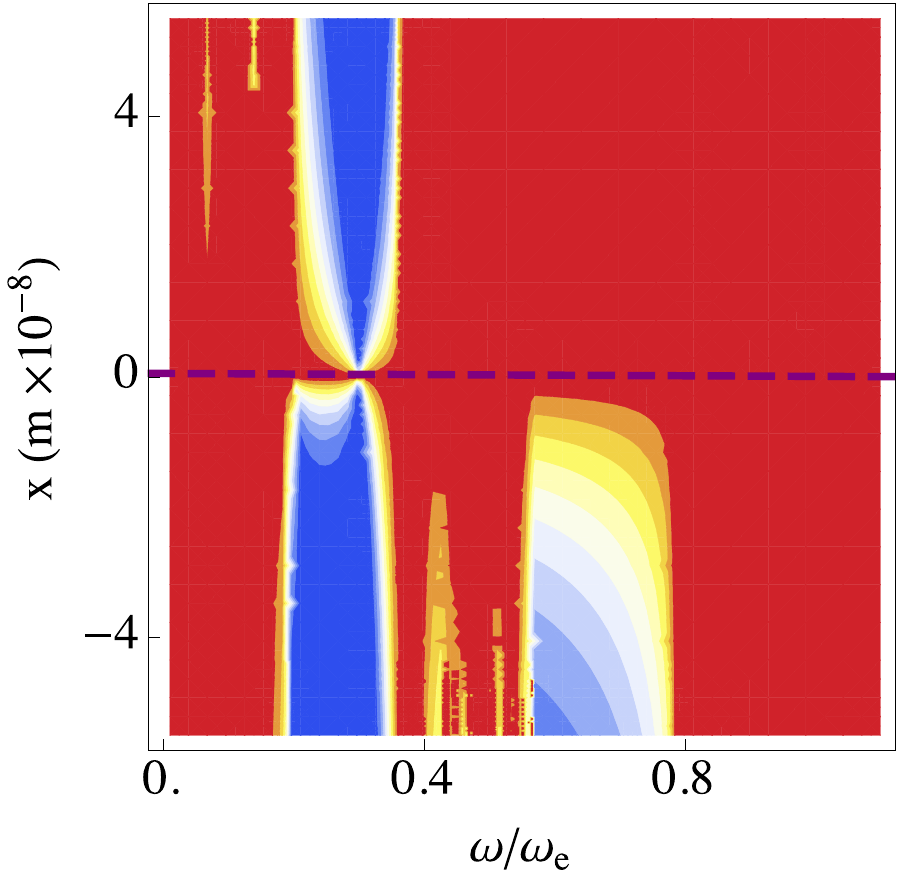}
\includegraphics[width=.042 \columnwidth]{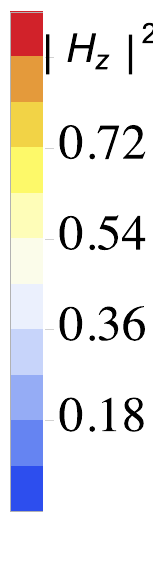}(c)\\
\includegraphics[width=.29 \columnwidth]{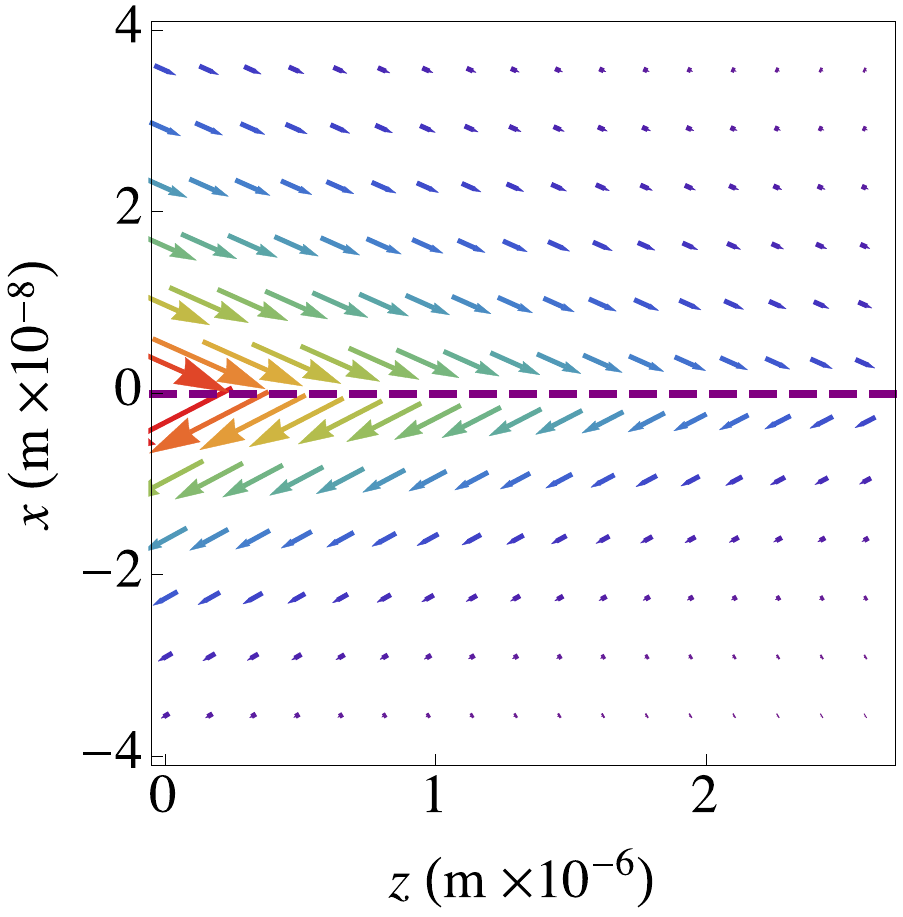}
\includegraphics[width=.042 \columnwidth]{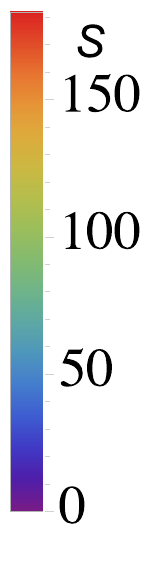}(d)
\includegraphics[width=.29 \columnwidth]{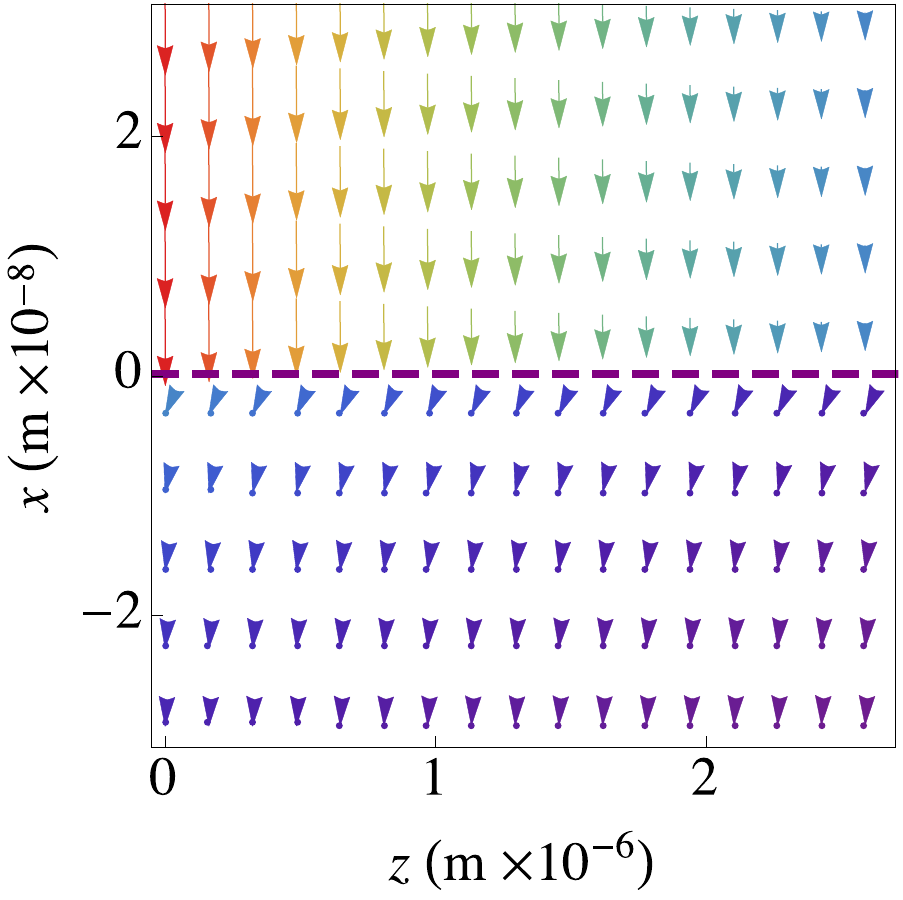}
\includegraphics[width=.042 \columnwidth]{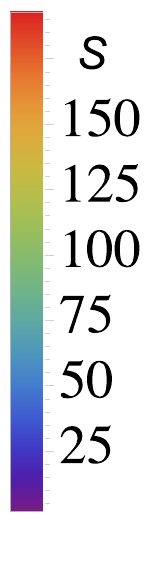}(e)
\includegraphics[width=.36 \columnwidth]{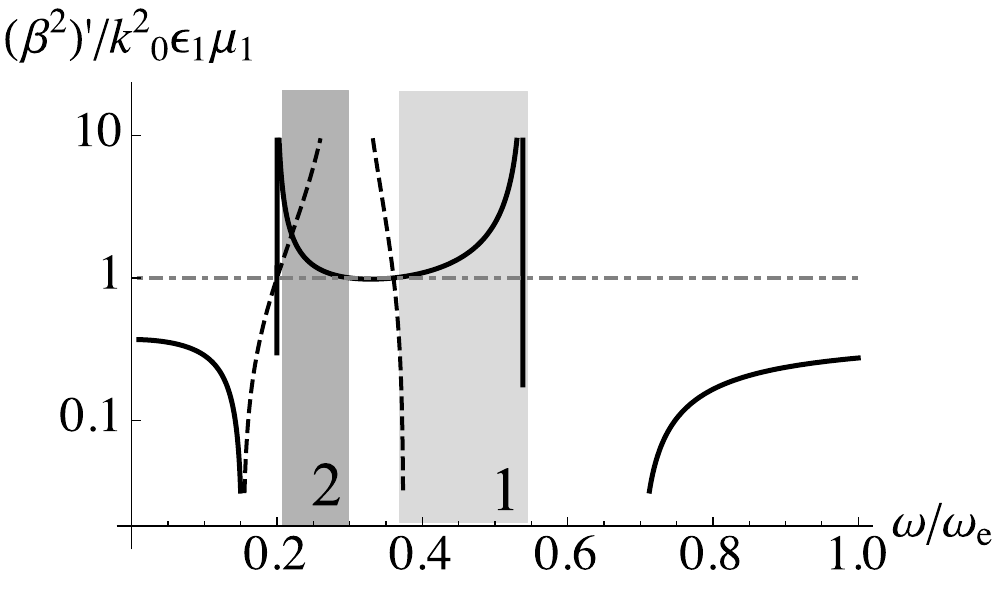}(f)
\includegraphics[width=.36 \columnwidth]{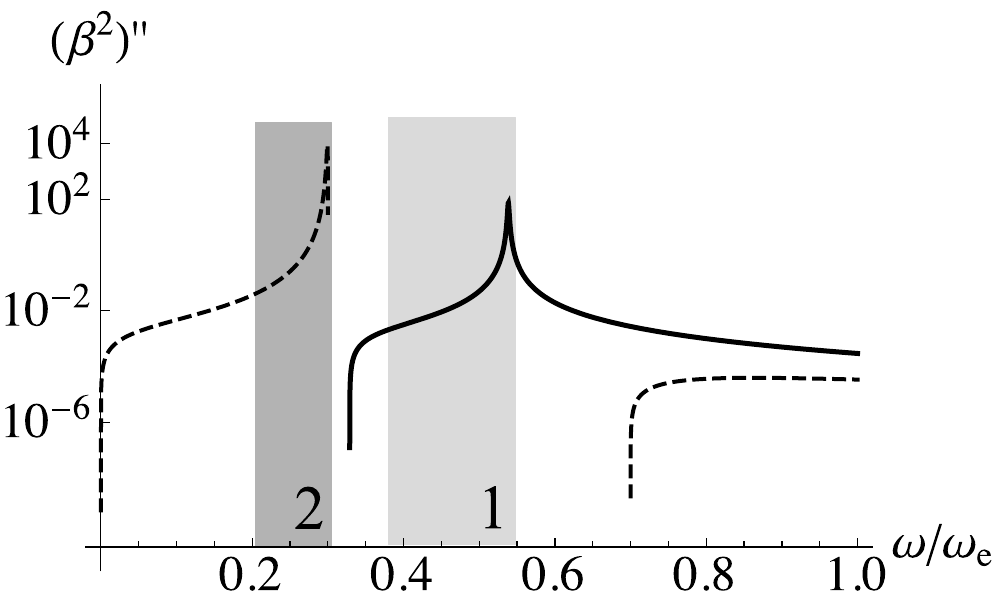}(g)
\caption{\label{fig:skindepth} 
Plots of (a) real part (solid line) and imaginary part (dashed line) of permittivity (black) and permeability (grey) of
dispersive LHI material,
(b)~electric field intensity
for TM SPPs and (c) magnetic field intensity 
for TE SPPs along the propagation direction $z$, as functions of distance from the interface $x$ and operating frequency.
Plots of time-averaged Poynting vector for TM SPPs at (d)~$\omega=0.5 \omega_{\mathrm e}$ when SPPs propagate and (e)~$\omega=0.71 \omega_{\mathrm e}$ when SPPs do not propagate.
Arrows show the direction and colours represent the magnitude of the Poynting vector as in plot legends.
Each vector arrowhead corresponds to a coordinate $(x,z)$.
The horizontal dashed line in plots (b)-(e)~represent the interface between dispersive and non-dispersive LHI materials.
(f)~and~(g)~are logarithmic plots of $\left(\beta^2\right)'/k^{2}_0 \varepsilon_1 \mu_1$ and $\left(\beta^2\right)''$ with respect to operating frequency, respectively.
The
dispersive LHI material parameters are 
        electric oscillation strength $F_{\mathrm{e}}=0.4$, 
	magnetic oscillation strength $F_{\mathrm{m}}=0.1$,
	electric-resonance frequency $\omega_{0_{\mathrm{e}}}=0.3 \omega_{\mathrm{e}}$,
	magnetic-resonance frequency $\omega_{0_{\mathrm{m}}}=0.2 \omega_{\mathrm{e}}$,
	$\omega_{\mathrm{e}}=1.37\times10^{16}~\mathrm{s}^{-1}$
	and
	EM damping $\Gamma_{\mathrm{m}}=\Gamma_{\mathrm{e}}=2.73\times10^{13}~\mathrm{s}^{-1}$~\cite{Lavoie2012}.
	The dielectric is treated as air with $\varepsilon_1=\mu_1=1$.
	Regions 1 and 2 are the propagating SPPs region for TM and TE SPPs, respectively, based on our characteristic equations in (\ref{Ca}).}
\end{figure}

The dispersive LHI material in our first example
has permittivity and permeability that can be
negative or positive in each case.
From Fig.~\ref{fig:skindepth}(a) we can see that the permeability
is negative for $0.2 \omega_{\mathrm{e}}<\omega<0.4 \omega_{\mathrm{e}}$ whereas, permittivity has negative values for $0.3 \omega_{\mathrm{e}}<\omega<\omega_{\mathrm{e}}$.
However, both permittivity and permeability have negative values for $0.3 \omega_{\mathrm{e}}<\omega<0.4 \omega_{\mathrm{e}}$.
Therefore, this example includes double-negative,~$\mu$-negative,~$\varepsilon$-negative and double-positive materials at the interface.

From our characteristic equations, we can specify the frequency region that SPPs can propagate at the interface.
Regions 1 and 2 in Fig.~\ref{fig:skindepth}(a) show the frequency regions where the conditions in Eq.~(\ref{Ca}) are satisfied for TM and TE SPPs, respectively.
For our choice of material parameters, TM SPPs propagate in the $\varepsilon$-negative frequency region, as well as in a small frequency window ($0.37 \omega_{\mathrm{e}}\lesssim\omega\lesssim0.4 \omega_{\mathrm{e}}$) where the material exhibits a double-negative refractive index. Conversely, TE SPPs propagate within the $\mu$-negative frequency window.

We investigate the behaviour of field intensity and Poynting vector of TM and TE SPPs at the interface to verify that the field intensity is concentrated at the interface and energy flux has components along the propagation direction to
verify the correctness of our characteristic equations.
The field intensity of TM and TE SPPs is shown in Figs.~\ref{fig:skindepth}(b)~and (c), respectively.
By comparing the field intensities with SPP-propagation regions in Figs.~\ref{fig:skindepth}(a) and (b), we can see that the frequency regions where the field intensity is concentrated near the interface (evanescent coupling), correspond to the frequencies at which we predict SPP propagation.

Figures~\ref{fig:skindepth}(d) and (e)~present the vector plots of the Poynting vector at the interface for two specific frequencies, where SPPs propagate and do not propagate, respectively.
For frequencies within the propagating SPPs regions, Poynting vectors have significant components along the interface; this specifies SPP propagation.
However, for frequencies outside the propagating SPPs regions, the Poynting vectors are nearly normal to the interface, indicating little to no energy flow along the interface.

The bounds on the real and imaginary parts of $\beta^2$ ((\ref{brlimit}) and (\ref{bilimit})) should hold in order to have propagating SPPs.
Figures~\ref{fig:skindepth}(f) and (g) present the behaviour of $\left(\beta^2\right)'/k^{2}_0 \varepsilon_1 \mu_1$ and $\left(\beta^2\right)''$ for TM and TE SPPs.
The logarithmic plots of Figs.~\ref{fig:skindepth}(f) and~(g)
just show positive values of $\left(\beta^2\right)'/k^{2}_0 \varepsilon_1 \mu_1$ and $\left(\beta^2\right)''$
whereas negative values are omitted in the plots.
These figures show that 
solutions satisfying our characteristic equations, also meet the physical definition of SPPs, (i.e., evanescent field confinement and energy propagation along the interface).
With the information about the generalized hypothetical material interface with air,
we investigate the realistic examples of the materials at the interface.

\subsection{Interfaces of Metamaterial, Electric Metal, and Magnetic Metal with Air}

Now we investigate the three cases of metamaterial, electric metal and effective magnetic metal interfaces with air, which simplify the expressions but capture the essential physics.
The behaviour of the permittivity and permeability of the three examples is shown in Figs.~\ref{fig:skindepthw0}(a), \ref{fig:skindepthe}(a) and \ref{fig:skindepthm}(a).
The corresponding field intensities
are shown in parts (b)~and (c) of these figures for TM and TE SPPs, respectively.
Poynting vectors at the interface are presented in parts (d) and (e)~
at two specific frequencies. The behaviour of the real and imaginary part of the squared propagation coefficient at the interfaces is presented in parts (f) and (g).
\begin{figure}[h]
\centering
\includegraphics[width=.36 \columnwidth]{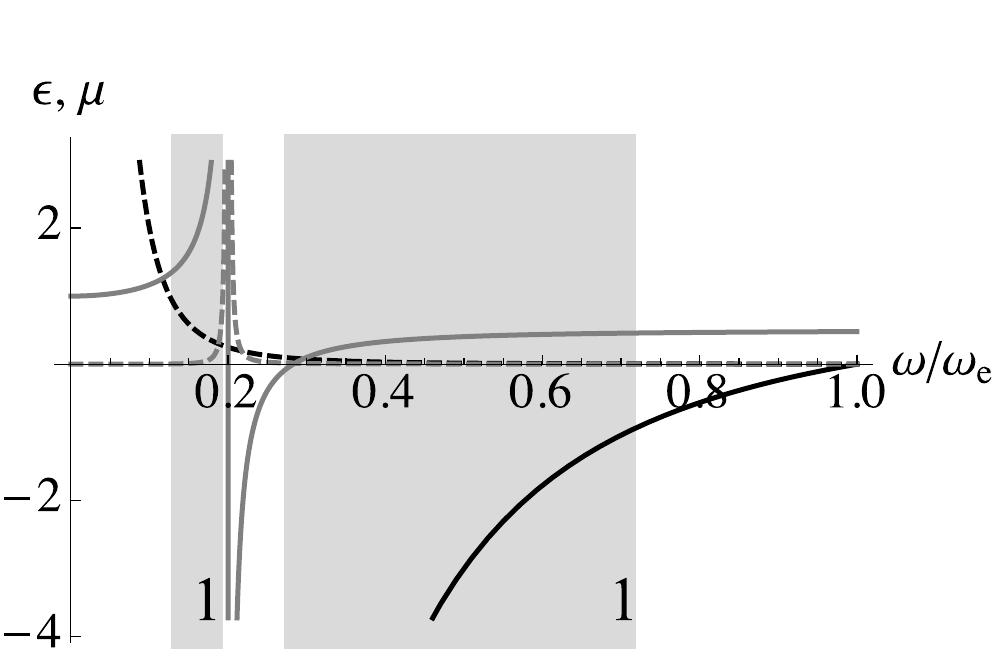}(a)\\
\includegraphics[width=.29 \columnwidth]{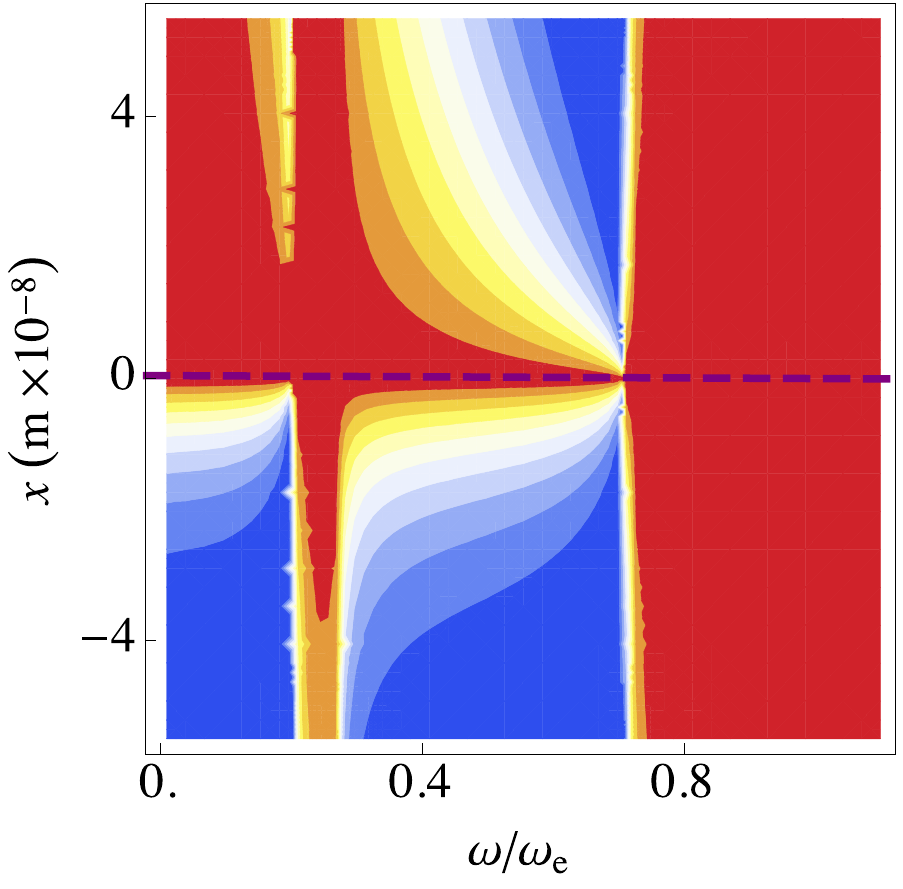}
\includegraphics[width=.042 \columnwidth]{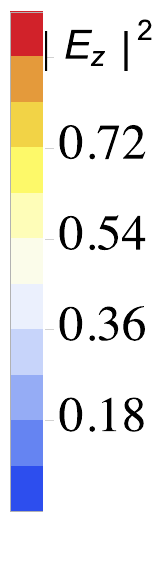}(b)
\includegraphics[width=.29 \columnwidth]{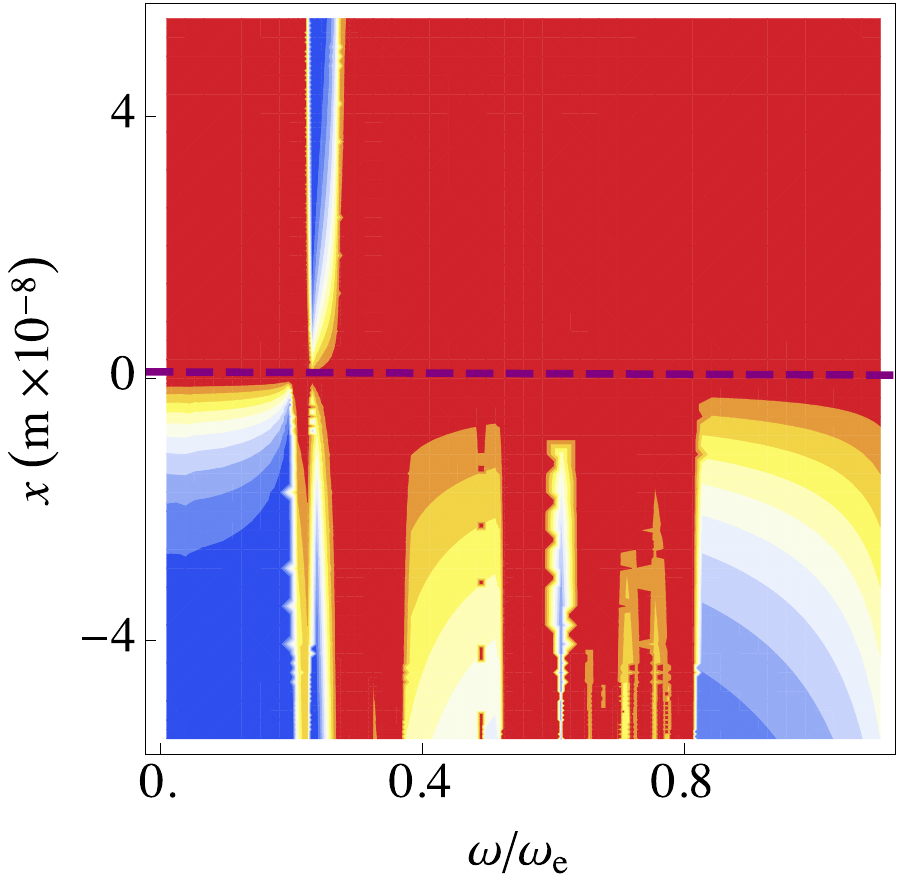}
\includegraphics[width=.042 \columnwidth]{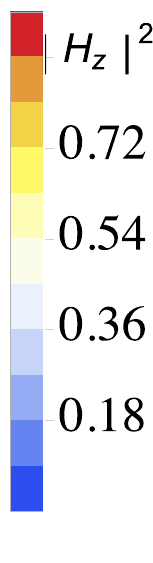}(c)\\
\includegraphics[width=.29 \columnwidth]{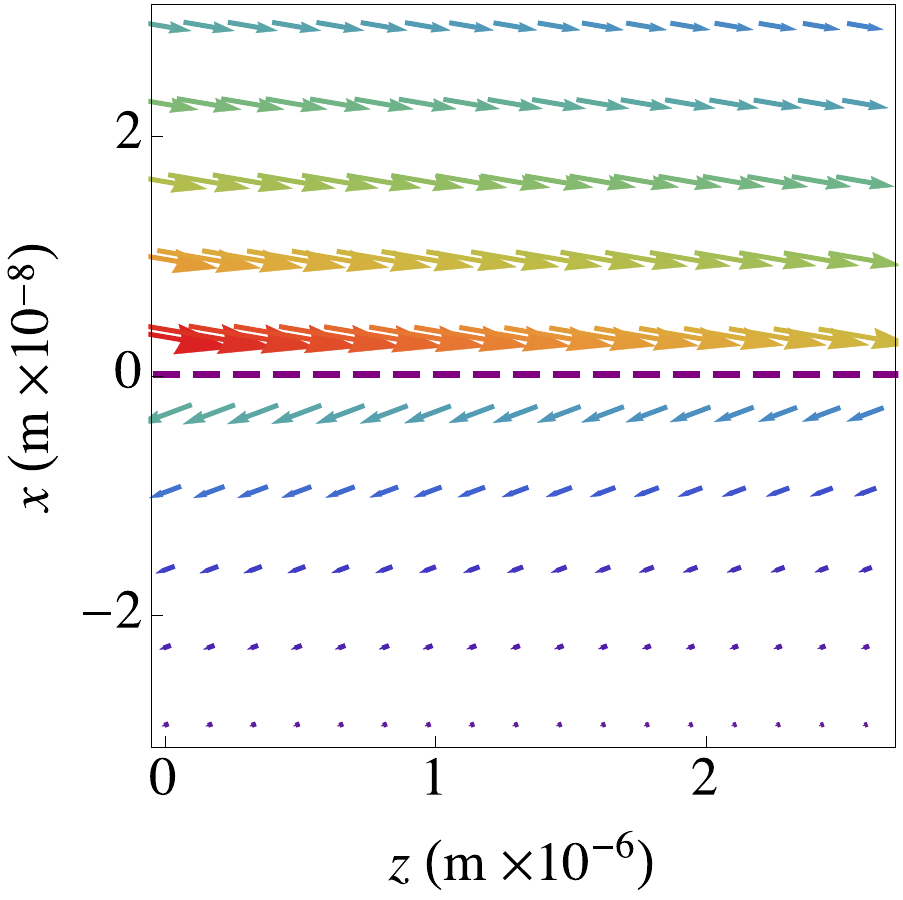}
\includegraphics[width=.042 \columnwidth]{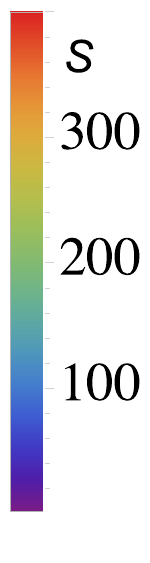}(d)
\includegraphics[width=.29 \columnwidth]{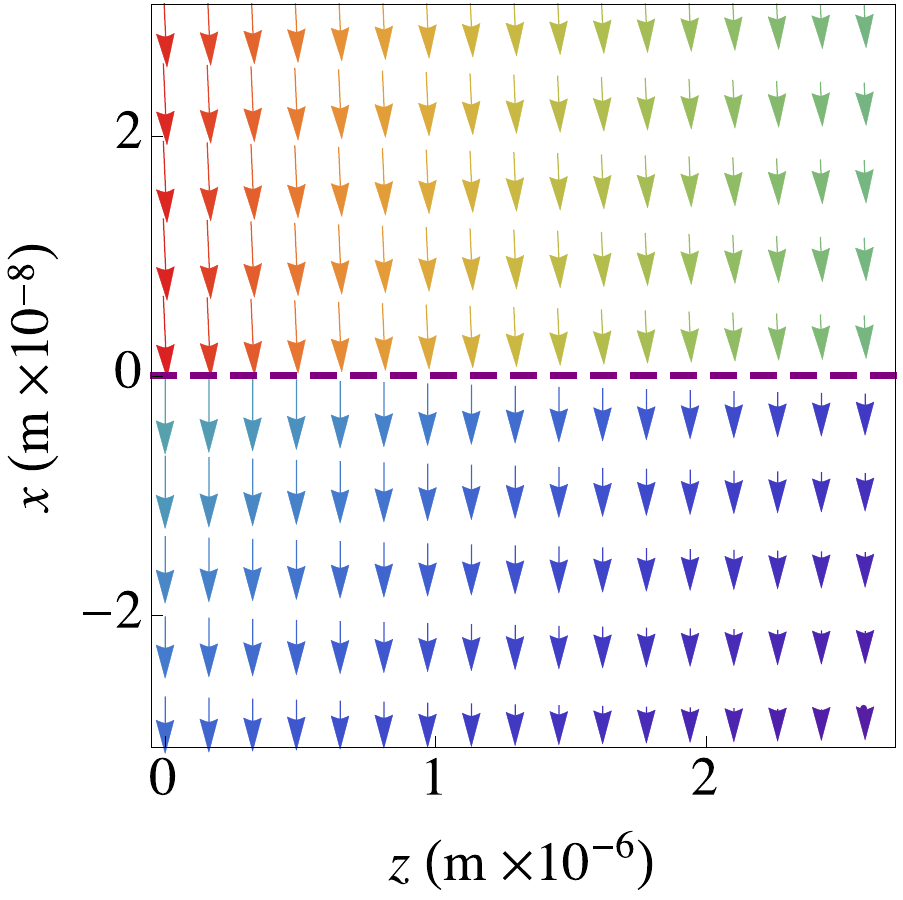}
\includegraphics[width=.042 \columnwidth]{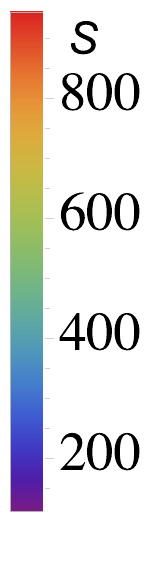}(e)
\includegraphics[width=.36 \columnwidth]{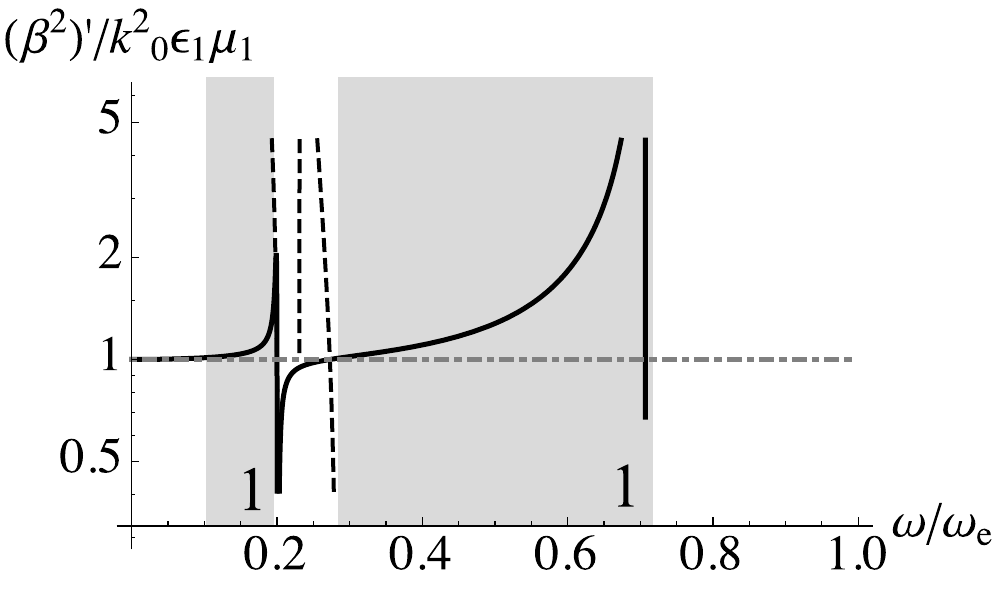}(f)
\includegraphics[width=.36 \columnwidth]{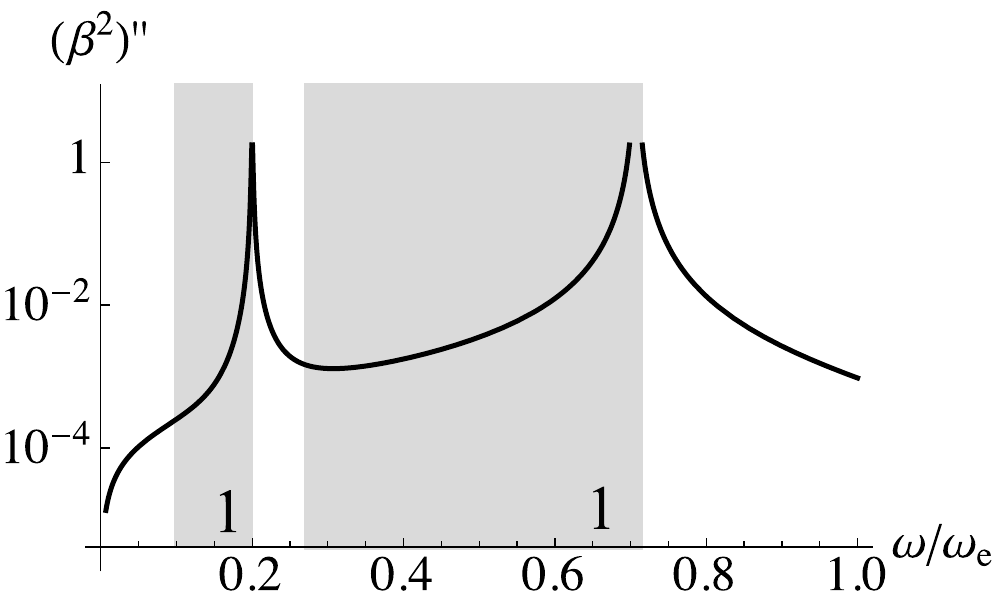}(g)
\caption{\label{fig:skindepthw0}
Plots of (a) real part (solid line) and imaginary part (dashed line) of permittivity (black) and permeability (grey) of
metamaterial, (b)~electric field intensity
for TM SPPs, (c) magnetic field intensity
for TE SPPs along the propagation direction $z$, (d) time-averaged Poynting vector for TM SPPs at $\omega=0.56 \omega_{\mathrm e}$ when SPPs propagate and (e)~time-averaged Poynting vector for TM SPPs at $\omega=0.21 \omega_{\mathrm e}$ when SPPs do not propagate.
Arrows show the direction of the Poynting vector and colours represent the magnitude of the Poynting vector as in plot legends.
The horizontal dashed line in plots (b)-(e)~represent the interface between dispersive and non-dispersive LHI materials.
(f)~and~(g) are logarithmic plots of $\left(\beta^2\right)'/k^{2}_0 \varepsilon_1 \mu_1$ and $\left(\beta^2\right)''$ with respect to operating frequency, respectively.
The metamaterial parameters are
as Fig.~\ref{fig:skindepth} with the exception $F_{\mathrm{e}}=1$, $F_\mathrm{m}=0.5$, $\omega_{0_{\mathrm{e}}}=0$ and $\omega_{0_{\mathrm{m}}}=0.2 \omega_{\mathrm{e}}$~\cite{Lavoie2012}.
Region 1 is TM propagating SPPs region, where there are no TE SPPs at this condition.
}
\end{figure}

For the metamaterial example, the negative permittivity and permeability frequency regions are $\omega<\omega_{\mathrm{e}}$ and $0.2 \omega_{\mathrm{e}}<\omega<0.3 \omega_{\mathrm{e}}$, respectively, as presented in Fig.~\ref{fig:skindepthw0}(a). The double-negative refractive-index region is at $0.2 \omega_{\mathrm{e}}<\omega<0.3 \omega_{\mathrm{e}}$.
Based on our characteristic equations, shown as shaded regions in Fig.~\ref{fig:skindepthw0}, no TE SPPs can propagate, and TM SPPs propagate in regions 1.
There is a small frequency region around the magnetic-resonance frequency where TE SPPs can propagate. AS this region is small, we do not shade it in Fig.~\ref{fig:skindepthw0}(a).

The behaviours of the field intensity and Poynting vector at the interface, Figs.~\ref{fig:skindepthw0}(b)-(e), confirm the accuracy of our characteristic equations
by showing that the field-intensity concentration of modes is localized to the surface and the Poynting vector has a large component along the propagation direction.
In Figs.~\ref{fig:skindepthw0}(f) and (g), for frequencies where $\left(\beta^2\right)'/k^{2}_0 \varepsilon_1 \mu_1>1$ and $\left(\beta^2\right)''>0$, SPPs propagate, which is in accordance with Figs.~\ref{fig:skindepthw0}(b)-(e).
For frequencies where $\left(\beta^2\right)'/k^{2}_0 \varepsilon_1 \mu_1<1$, $\left(\beta^2\right)'/k^{2}_0 \varepsilon_1 \mu_1\approx1$ and where $\left(\beta^2\right)''<0$, SPPs do not propagate.

In the dual condition of Fig.~\ref{fig:skindepthw0}, where $\omega_{{0}_{\mathrm{m}}}=0$, no TM SPPs can propagate and TE SPPs propagate at the interface.
Based on our characteristic equations, TE and TM SPPs cannot be supported at negative-index metamaterial interfaces with arbitrary values of permittivity and permeability.
\begin{figure}
\centering
\includegraphics[width=.36 \columnwidth]{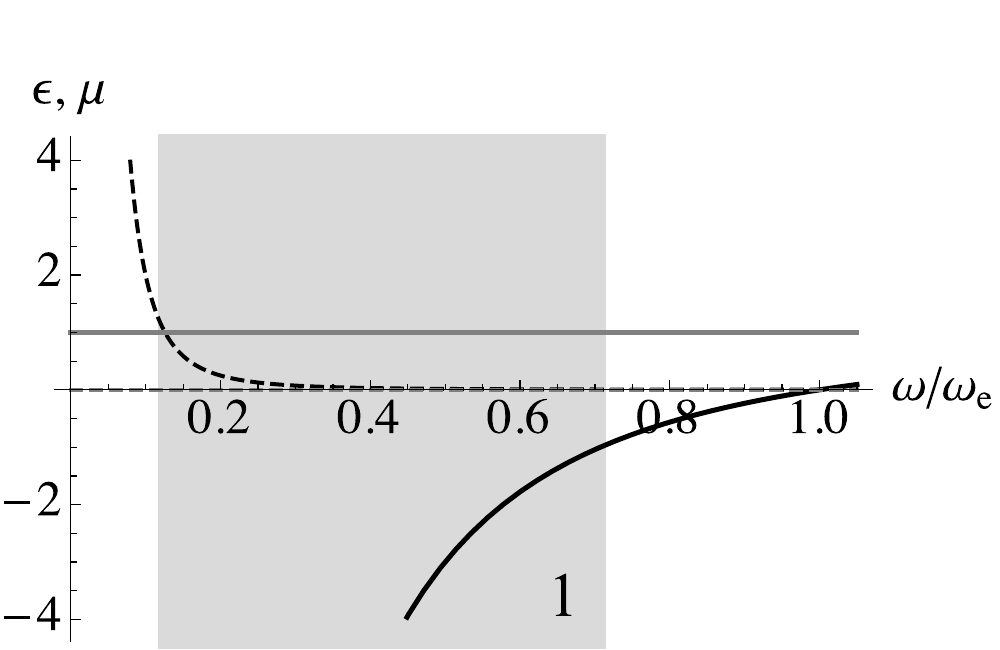}(a)\\
\includegraphics[width=.29 \columnwidth]{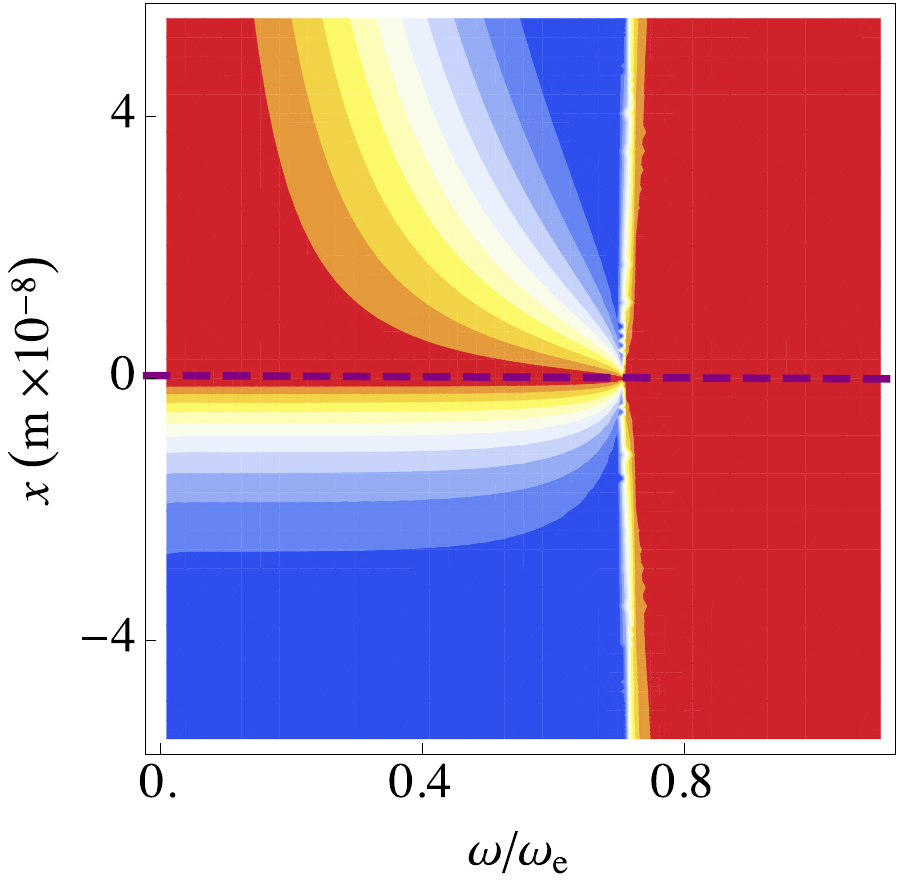}
\includegraphics[width=.042 \columnwidth]{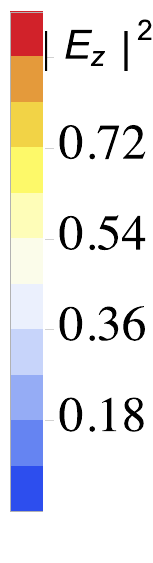}(b)
\includegraphics[width=.29 \columnwidth]{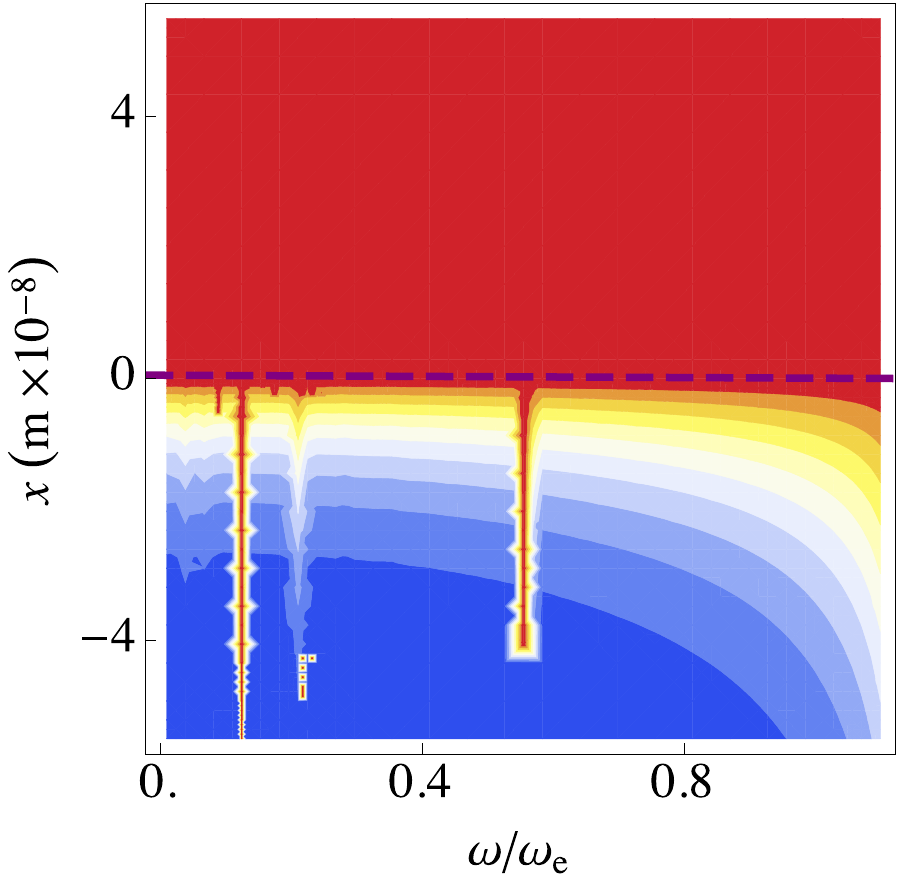}
\includegraphics[width=.042 \columnwidth]{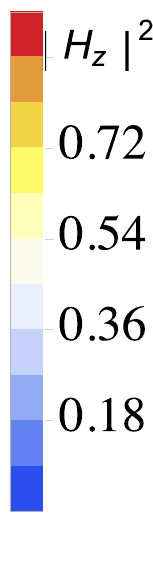}(c)\\
\includegraphics[width=.29 \columnwidth]{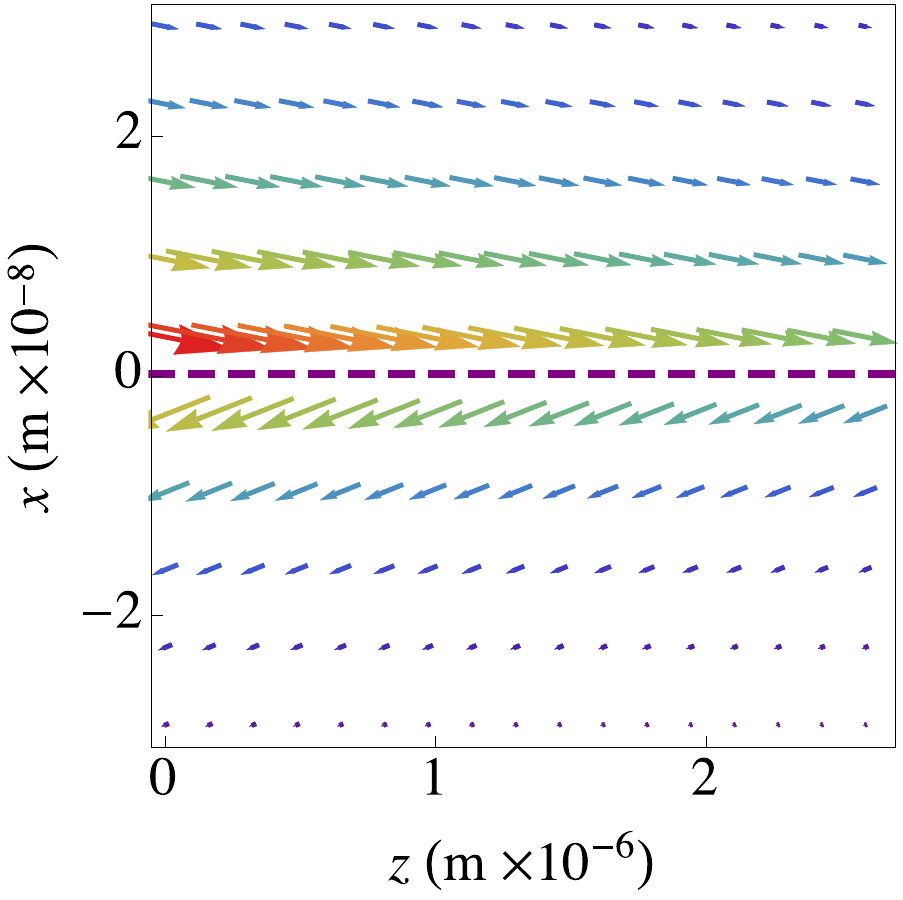}
\includegraphics[width=.042 \columnwidth]{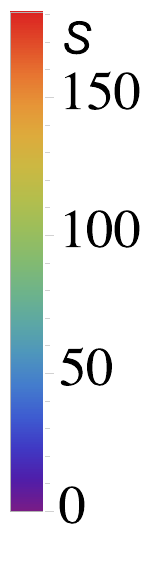}(d)
\includegraphics[width=.29 \columnwidth]{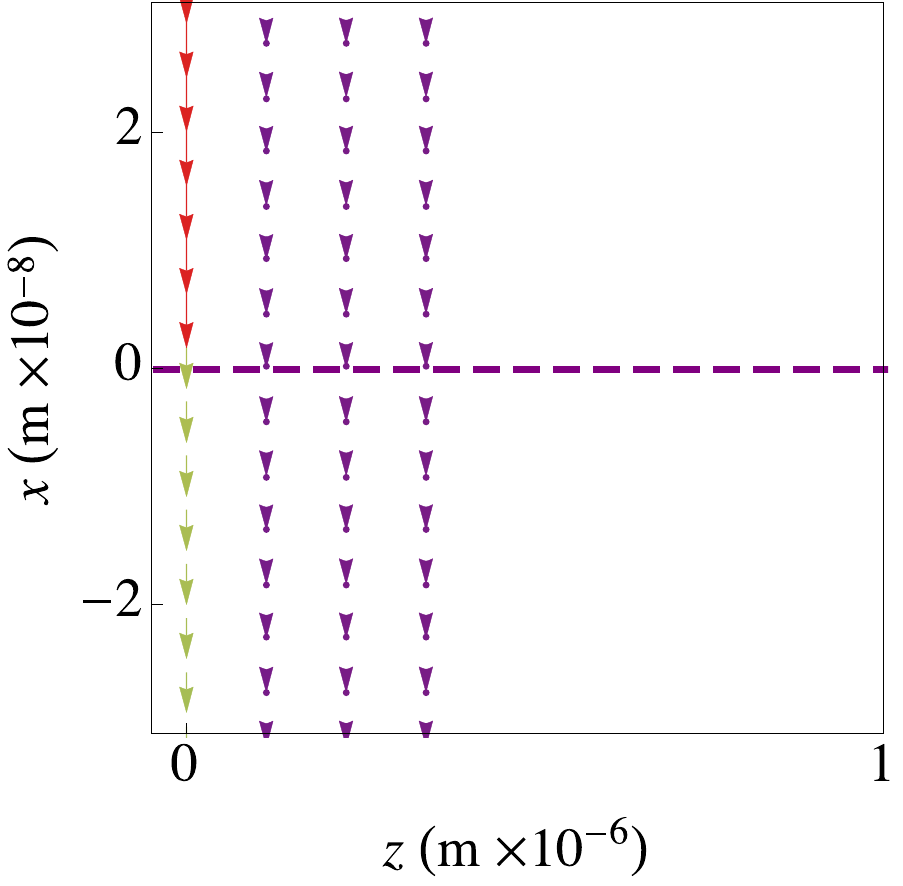}
\includegraphics[width=.042 \columnwidth]{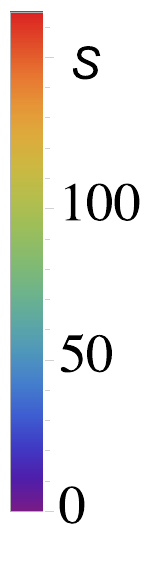}(e)
\includegraphics[width=.36 \columnwidth]{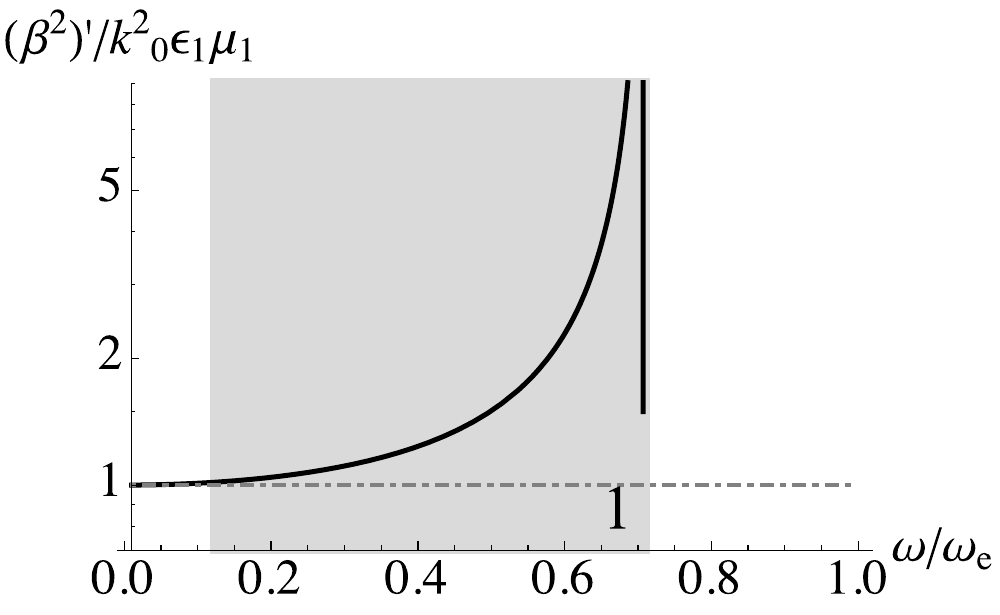}(f)
\includegraphics[width=.36 \columnwidth]{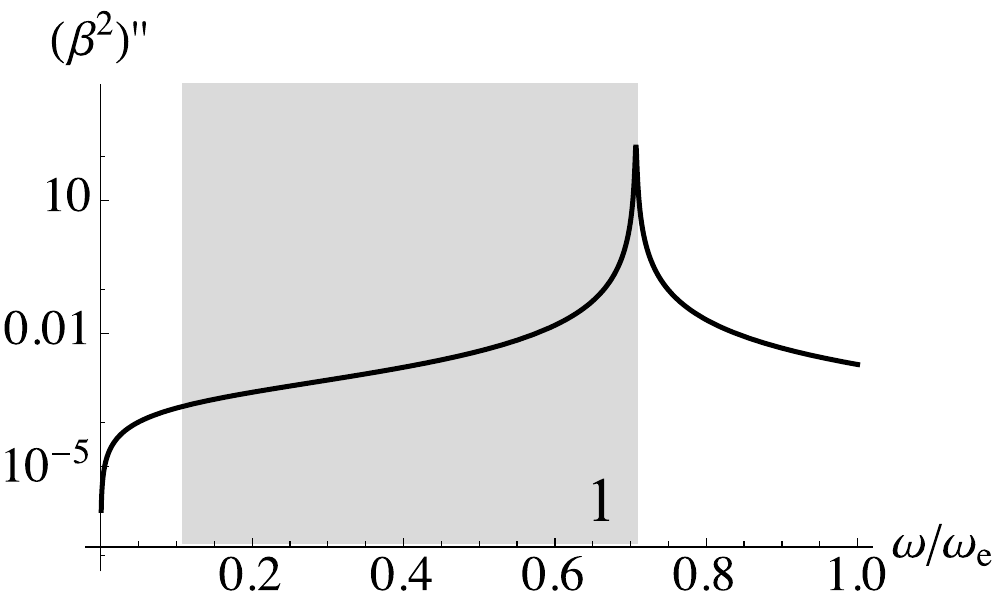}(g)
\caption{\label{fig:skindepthe}
Plots of (a) real part (solid line) and imaginary part (dashed line) of permittivity (black) and permeability (grey) of
electric metal, (b)~electric field intensity
for TM SPPs, (c) magnetic field intensity
for TE SPPs, (d) time-averaged Poynting vector for TM SPPs at $\omega=0.6 \omega_{\mathrm e}$ when SPPs propagate and (e)~time-averaged Poynting vector for TM SPPs at $\omega=0.9 \omega_{\mathrm e}$ when SPPs do not propagate.
Arrows show the direction of Poynting vector and the colours represent the magnitude of the Poynting vector as in plot legends.
The horizontal dashed line in plots (b)-(e)~represent the interface between dispersive and non-dispersive LHI materials.
(f)~and~(g) are logarithmic plots of $\left(\beta^2\right)'/k^{2}_0 \varepsilon_1 \mu_1$ and $\left(\beta^2\right)''$ with respect to operating frequency, respectively.
The parameters are as Fig.~\ref{fig:skindepthw0} with the exception $\mu_2=1$.
Region 1 is the TM propagating SPPs region, where there are no TE SPPs at this condition.
}
\end{figure}

\begin{figure}
\centering
\includegraphics[width=.36 \columnwidth]{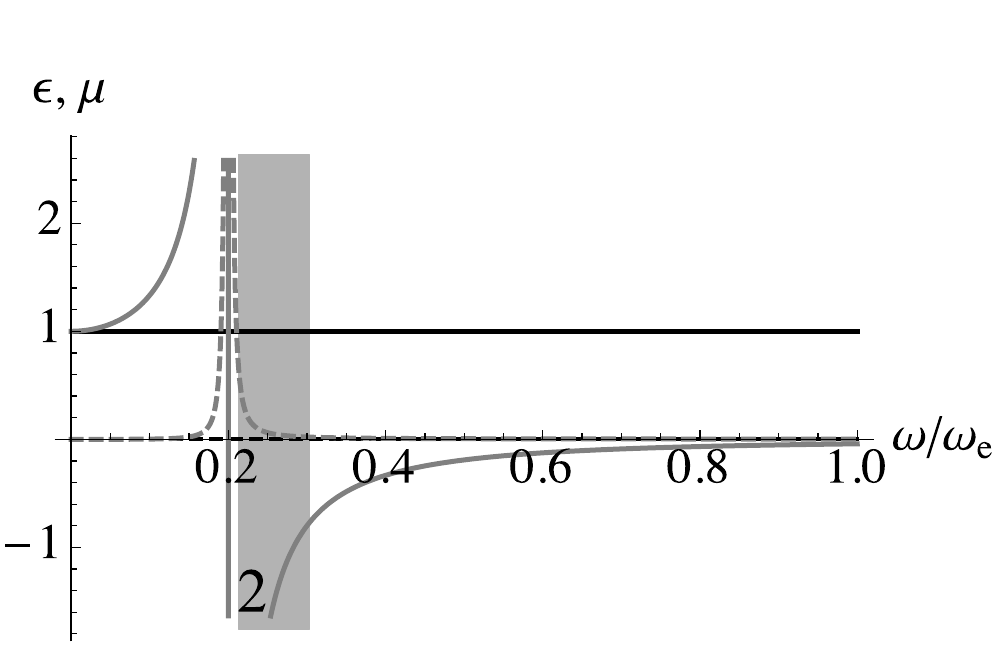}(a)\\
\includegraphics[width=.29 \columnwidth]{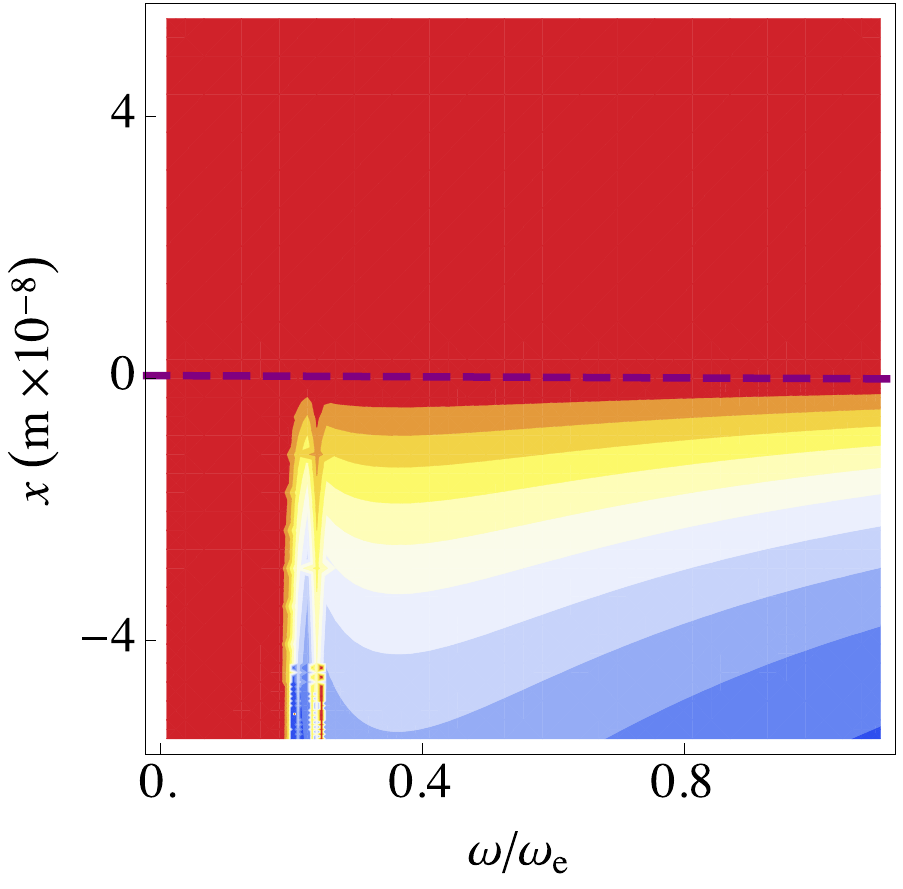}
\includegraphics[width=.042 \columnwidth]{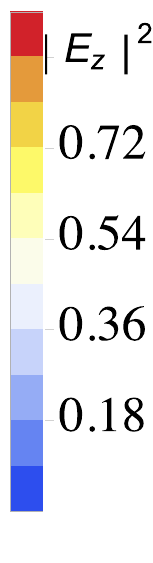}(b)
\includegraphics[width=.29 \columnwidth]{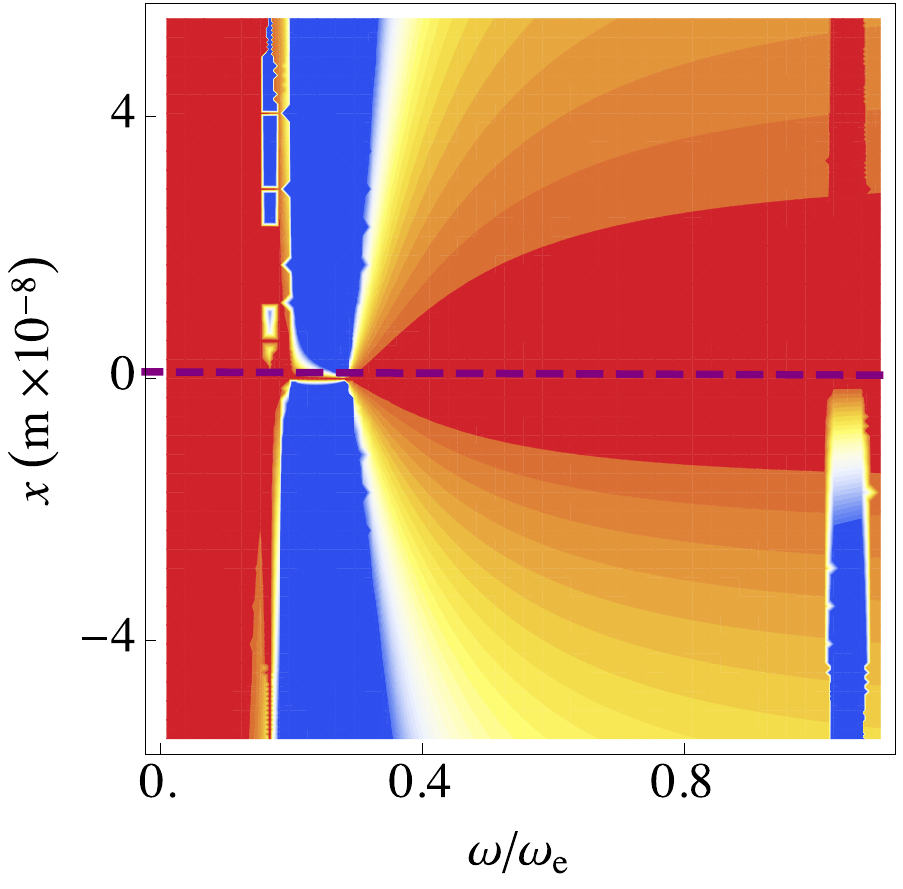}
\includegraphics[width=.042 \columnwidth]{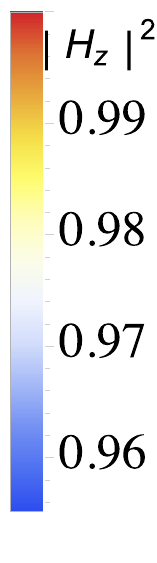}(c)\\
\includegraphics[width=.29 \columnwidth]{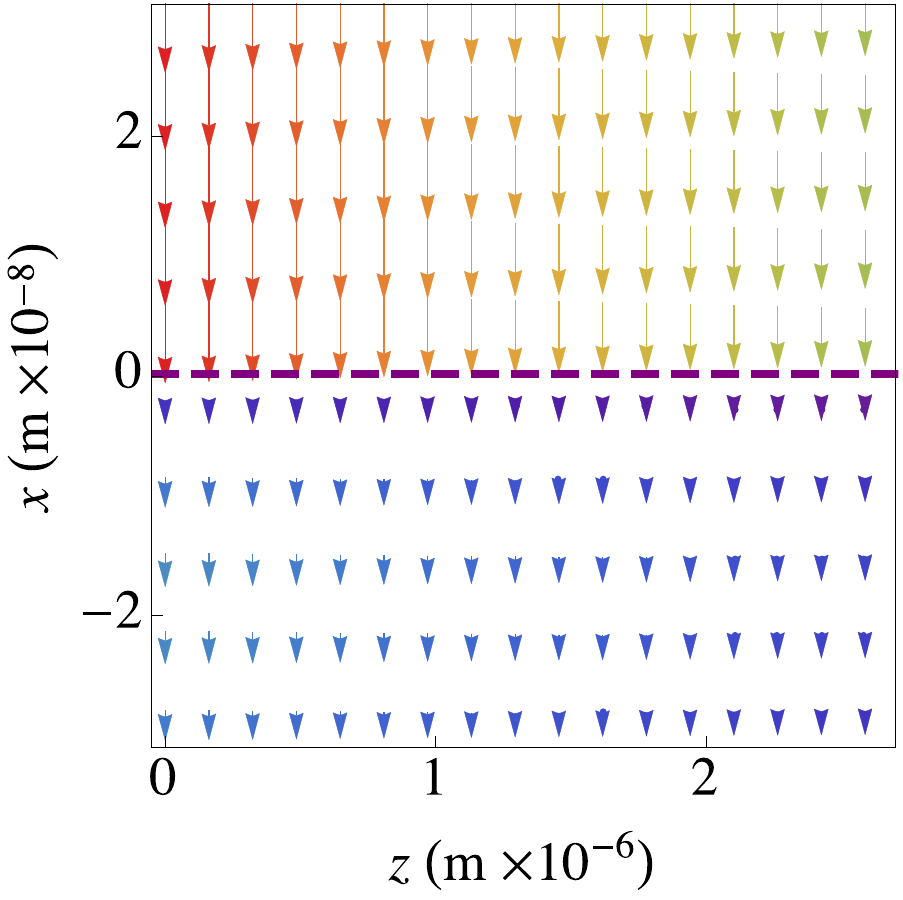}
\includegraphics[width=.042 \columnwidth]{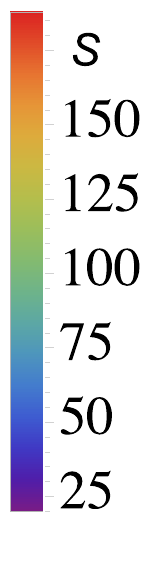}(d)
\includegraphics[width=.29 \columnwidth]{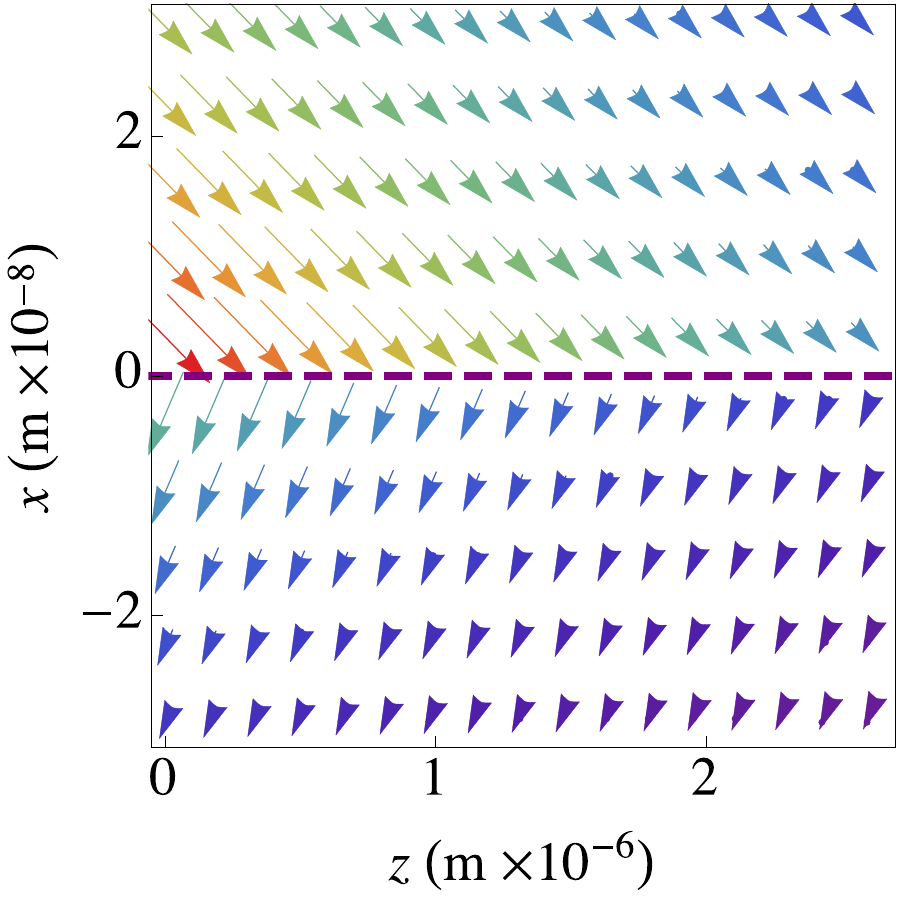}
\includegraphics[width=.042 \columnwidth]{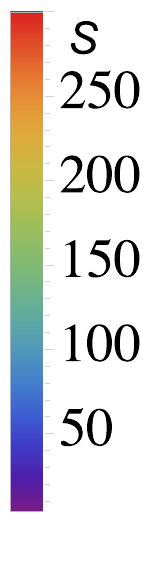}(e)
\includegraphics[width=.36 \columnwidth]{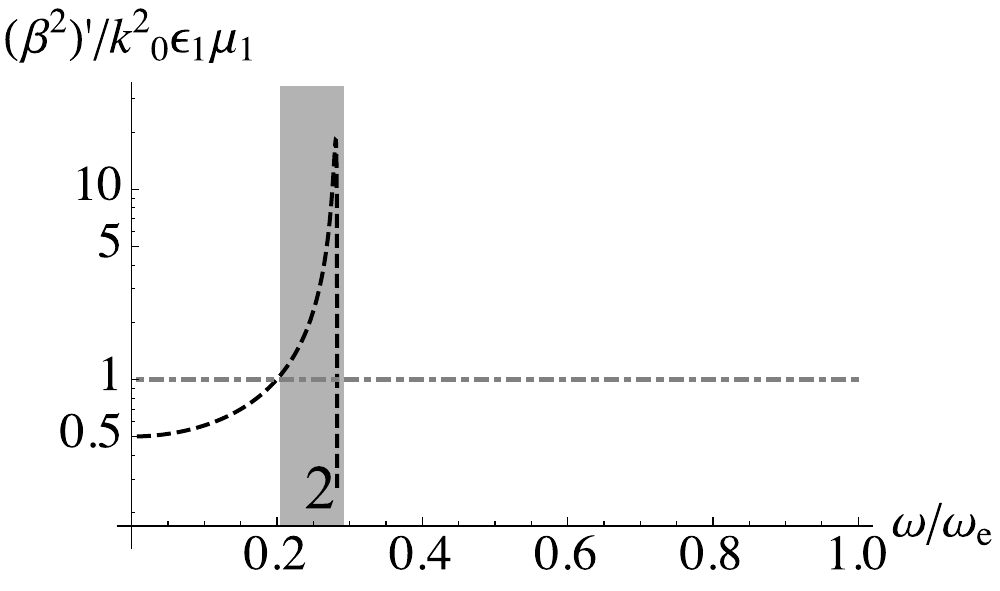}(f)
\includegraphics[width=.36 \columnwidth]{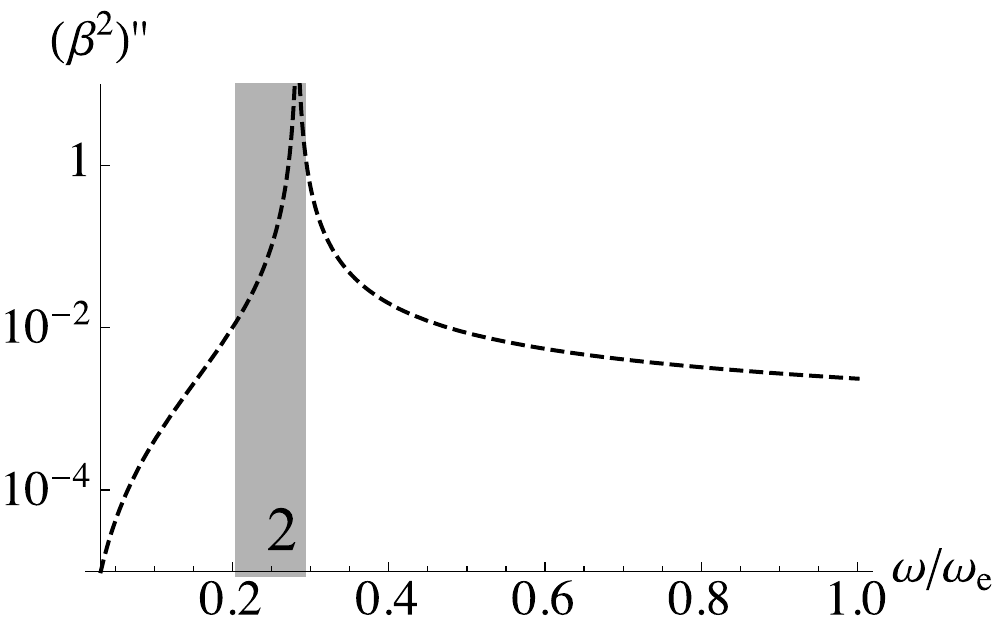}(g)
\caption{\label{fig:skindepthm} 
Plots of (a) real part (solid line) and imaginary part (dashed line) of permittivity (black) and permeability (grey) of
dispersive LHI material, (b)~electric field intensity
for TM SPPs, (c) magnetic field intensity
for TE SPPs, (d)
time-averaged Poynting vector for TM SPPs at $\omega=0.1 \omega_{\mathrm e}$ when SPPs do not propagate,
and (e)~time-averaged Poynting vector for TE SPPs at $\omega=0.24 \omega_{\mathrm e}$ when SPPs propagate.
Arrows show the direction of Poynting vector and the colours represent the magnitude of the Poynting vector as in plot legends.
The horizontal dashed line in plots (b)-(e)~represent the interface between dispersive and non-dispersive LHI materials.
(f)~and~(g) are logarithmic plots of $\left(\beta^2\right)'/k^{2}_0 \varepsilon_1 \mu_1$ and $\left(\beta^2\right)''$ with respect to operating frequency, respectively.
The plots parameters and symbol descriptions are as Fig.~\ref{fig:skindepth}.
The parameters are as Fig.~\ref{fig:skindepthw0} with the exception $\varepsilon_2=1$ and $F=1$.
Region 2 is the TE propagating SPPs region, where there are no TM SPPs at this condition.
}
\end{figure}

The examples of electric and effective magnetic metals interfaces with air are presented in Figs.~\ref{fig:skindepthe} and \ref{fig:skindepthm}, respectively.
In the case of an electric metal interface with air, where permittivity changes sign at the interface, presented in Fig.~\ref{fig:skindepthe}(a), just TM SPPs propagate at the interface.
The reverse is true at the interface of the effective magnetic metal with air where just TE SPPs propagate, as presented in Fig.~\ref{fig:skindepthm}(a).
In the propagating SPPs regions 1 and 2 in Figs.~\ref{fig:skindepthe} and \ref{fig:skindepthm}, field intensity and Poynting vector verify SPP propagation
by showing the localization of intensity concentration of modes to the surface and the large component of the Poynting vector along the propagation direction
and $\left(\beta^2\right)'/k^{2}_0 \varepsilon_1 \mu_1>1$ and $\left(\beta^2\right)''/k_0>0$ at these regions.

To sum up, we investigate the behaviour of SPPs at the interfaces of lossy dispersive LHI materials with lossless nondispersive LHI materials.
We present examples of lossy dispersive hypothetical material, metamaterial, electric metal, and effective magnetic metal planar interfaces with air and investigate the behaviour of SPPs at these interfaces. 
By investigating the field intensity and the Poynting vector at the interface we verify the viability of our characteristic equations.

\section{Discussion}
\label{sec:discussion}

The characteristic equations specify the regions that SPPs propagate along the interface.
To verify the consistency of our characteristic equations,
we investigate the behaviour of field intensity concentration
and the Poynting vector at the interface.
Our investigations verify that
at regions where our characteristic equations show SPP propagation,
field intensity concentration is localized to the interface
and the Poynting vector has large components along the propagation direction,
compared to the condition that SPPs do not propagate.

At regions for which the characteristic equations predict no SPPs propagation,
then only weak coupling, or else no coupling at all,
occurs between the EM field
and coherent free-charge oscillation at the interface.
This behaviour leads to large losses, due to energy leaking into the bulk material. 
Thus, the fields are decoupled from the interface
and do not display the evanescent decay that is characteristic of propagating SPP modes.
At frequency regions where the characteristic equations verify SPP propagation,
field intensity concentration is localized at the interface.

The Poynting vector has components parallel and perpendicular to the propagation direction,
where the parallel component changes direction at the interface and the perpendicular component
is directed toward the lossy material.
The parallel component of the Poynting vector at the interface
signifies SPP propagation,
and the perpendicular
component
corresponds to energy leaking from the interface toward the lossy material.
When the parallel component of the Poynting vector is large,
compared to the perpendicular component,
energy is predominantly travelling
along the interface.
This component of energy is large
compared to the portion that leaks away from the interface,
and SPPs propagate.

At lossless interfaces,
the Poynting vector is completely parallel to the interface,
as the components of electric field along the propagation direction
and magnetic field perpendicular to the propagation direction
are 90 degrees out of phase
and their cross-product,
which corresponds to the component of the Poynting vector
perpendicular to the interface, is zero.
However, at interfaces between lossy materials, the Poynting vector has a component perpendicular
to the interface.

We have employed the characteristic equations to
determine regions for which TE and TM SPPs
propagate along the material interfaces with a dielectric.
TM and TE SPPs
can propagate at the interface of the hypothetical LHI materials with air.
As an example, we have shown that
a metamaterial interface with air supports TM SPPs
but not TE SPPs.
Only at the magnetic-resonance frequency
where the magnetic permeability is large,
the propagation coefficient of TE SPPs assumes a high value
and SPPs propagate at the interface.
However, at this resonance frequency,
the loss rate is high, causing TE SPPs to dissipate into the bulk
almost immediately after excitation.
Electric metals facilitate the propagation of TM SPPs at the interface with air
while
TE SPPs propagate at the interface of magnetic metals with air.

Our characteristic equations work
equally well for both forward-propagating ($\beta'>0$ and $\beta''>0$)
and backward-propagating ($\beta'<0$ and $\beta''<0$) SPPs at lossy interfaces.
Deriving bounds on
real and imaginary parts of $\beta^2$
are key steps towards deriving the SPP characteristic equation.
At regions where $(\beta^2)'/k^2_0 \varepsilon_1 \mu_1>1$ and $(\beta^2)''>0$,
SPPs propagate along the interface.
We establish these bounds on $\beta^2$
by applying the conditions on the wavenumber at lossy interfaces
and considering $\beta'$ and $\beta''$ having positive values
for forward propagating SPPs at lossy interfaces.
The negative value of $\beta''$ when $\beta'$ is positive means gain
that is not acceptable for lossy interfaces.
However, 
The negative values of both $\beta'$ and $\beta''$ together is acceptable,
as this condition on $\beta'$ and $\beta''$ means the wave is either backward or the source propagation direction has changed.

In summary our bounds on complex squared propagation coefficients,
coupled with the wavenumber conditions at lossy interfaces,
yields the characteristic equations.
These characteristic equations reveal when SPP propagation can or cannot occur for both TE and TM SPPs.
These SPP propagation solutions are obtained as mathematical results,
which we confirm by checking against physical intuition through analyzing field-intensity and Poynting-vector properties.

\section{Conclusions}
\label{sec:conclusions}
We have extended the case of SPPs at lossless metallic interfaces to define and derive characteristic equations for SPPs at interfaces between lossless materials and those with arbitrary material properties, including loss.
To derive the characteristic equations we introduced strict bounds on the
real and imaginary parts of the squared propagation coefficient to decide whether a given mode is a propagating SPP or not.
Our results,
which include the quite general case of LHI materials meeting at planar interfaces,
include all the previous results in literature as special cases and corrects previous errors due to ignoring losses.

We have confirmed the consistency of our characteristic equations by showing that the intensity concentration of modes is localized to the surface for the characteristic equations satisfying the bounds and the Poynting vector has a large component along the propagation direction compared to the component perpendicular to the surface.
Ascertaining whether SPPs can exist or not at the interfaces is a valuable and easy first step, using our characteristic equation, to decide whether a given study or application could be viable.
Our techniques could be valuable for designing plasmonic circuits and waveguides.

\section*{Acknowledgments}
BCS acknowledges financial support from NSERC, Alberta Innovates,
China's
1000 Talent Plan,
NSFC (Grant No.\ GG2340000241)
and the Institute for Quantum Information and Matter,
which is an NSF Physics Frontiers Center (NSF Grant PHY-1125565)
with support of the Gordon and Betty Moore Foundation (GBMF-2644).
BRL acknowledges financial support from NSERC CREATE I3T Program.

\section*{References}
\bibliography{SPPReferences}

\end{document}